\documentclass[prl,reprint,notitlepage,groupedaddress,floatfix,superscriptaddress]{revtex4-2}
\usepackage[english]{babel}
\usepackage[T1]{fontenc}
\usepackage[final]{graphicx}
\usepackage{natbib}
\usepackage{amssymb}
\usepackage{amsmath}
\usepackage{mathrsfs}
\usepackage{xcolor}
\usepackage{bm}
\usepackage{siunitx}
\usepackage{braket}
\usepackage{dcolumn}   
\usepackage[normalem]{ulem}
\usepackage{epstopdf}
\usepackage{lmodern}

\newcommand{\ak}[1]{{\color{black}{#1}}}

\newcommand{\dd}[1]{{\color{black}{#1}}}

\usepackage{physics}
\usepackage{scalerel,stackengine}

\makeatletter
\begin{document}
\title{Quantum beats of a macroscopic polariton condensate in real space}

\author{R.V. Cherbunin}
\affiliation{Department of Physics, St. Petersburg State University, University Embankment, 7/9, St. Petersburg, 199034, Russia}
\affiliation{Russian Quantum Center, Skolkovo, Moscow, 121205, Russia}
\email{r.cherbunin@spbu.ru}
\author{A. Liubomirov}
\affiliation{Department of Physics, St. Petersburg State University, University Embankment, 7/9, St. Petersburg, 199034, Russia}
\affiliation{Russian Quantum Center, Skolkovo, Moscow, 121205, Russia}
\author{D. Novokreschenov}
\affiliation{Russian Quantum Center, Skolkovo, Moscow, 121205, Russia}
\affiliation{Abrikosov Center for Theoretical Physics, MIPT, Dolgoprudnyi, Moscow Region 141701, Russia}
\author{A. Kudlis}
\affiliation{Russian Quantum Center, Skolkovo, Moscow, 121205, Russia}
\affiliation{Abrikosov Center for Theoretical Physics, MIPT, Dolgoprudnyi, Moscow Region 141701, Russia}
\email{andrewkudlis@gmail.com}
\author{A.V. Kavokin}
\affiliation{Russian Quantum Center, Skolkovo, Moscow, 121205, Russia}
\affiliation{Abrikosov Center for Theoretical Physics, MIPT, Dolgoprudnyi, Moscow Region 141701, Russia}
\affiliation{International Center for Polaritonics, Westlake university, Shilongshan Road, 18, Hangzhou, 310024, China}
\email{kavokinalexey@gmail.com}

\date{\today}

\begin{abstract}
We experimentally observe harmonic oscillations in a bosonic condensate of exciton-polaritons confined within an elliptical trap. These oscillations arise from quantum beats between two size-quantized states of the condensate, split in energy due to the trap's ellipticity. By precisely targeting specific spots inside the trap with nonresonant laser pulses, we control frequency, amplitude, and phase of these quantum beats. The condensate wave function dynamics is visualized on a streak camera and mapped to the Bloch sphere, demonstrating Hadamard and Pauli-Z operations. We conclude that a qubit based on a superposition of these two polariton states would exhibit a coherence time exceeding the lifetime of an individual exciton-polariton by at least two orders of magnitude.
\end{abstract} 

\maketitle

\textit{Introduction.---} Quantum beats are a widely spread phenomenon that is characteristic of the coherent dynamics of a two-level quantum system excited in a superposition state~\cite{PhysRevA.35.1720,J_N_Dodd_1964,PhysRevLett.30.948}. The period of the beats is determined by the energy splitting of two participating quantum states~\cite{J_N_Dodd_1964}, while their decay time characterizes the decoherence processes that are necessarily present in any quantum system coupled to an environment~\cite{PhysRevLett.30.948,shah2013ultrafast}. In condensed matter physics, quantum beats caused by size quantization of electron~\cite{PhysRevLett.68.2216}, hole~\cite{PhysRevB.46.10460} or exciton~\cite{PhysRevLett.64.1801} wave functions, Zeeman splitting~\cite{PhysRevLett.66.2491}, spin-orbit~\cite{PhysRevLett.122.147401}, hyperfine~\cite{PhysRevB.94.081201} interactions $etc$ have been documented. Quantum beats are typically detected by optical methods, including time-resolved photoluminescence~\cite{alexandrov1964interference}, pump-probe~\cite{PhysRevA.35.1720}, four-wave mixing~\cite{PhysRevA.48.R1765}, Faraday-rotation~\cite{PhysRevB.80.104436} spectroscopy, spin noise~\cite{MULLER2010569} and photon-echo~\cite{PhysRevLett.13.567} spectroscopies as well as several other techniques. These approaches enable detection of the oscillations of intensity, polarisation, coherence degree and/or other measurable characteristics of light caused by quantum beats in crystals, molecules and other matter objects coupled to light~\cite{Scully1997-aq}. 
In this context, bosonic light-matter quasiparticles, namely, exciton-polaritons~\cite{Deng2002,Kasprzak2006,RevModPhys.85.299} represent a unique laboratory for studies of the optical manifestations of quantum beats. In the strong exciton-photon coupling regime in semiconductor microcavities, the quantum beats between exciton-polariton eigen-modes frequently referred to as Rabi-oscillations have been observed since mid-1990s~\cite{norris1995m,10.1063/1.113827}. \ak{The true quantum Rabi-oscillations were demonstrated by means of the solid-state cavity QED~\cite{Hennessy2007}}. Next, a number of experiments demonstrated the complex coherent oscillatory dynamics of the polarization and intensity of light emitted by linear combinations of various polariton states~\cite{PhysRevResearch.3.013007,PhysRevB.72.075317} that e.g. led to generation of full Bloch light beams~\cite{PhysRevResearch.3.013007}. A fundamental factor that limits the decay time of polariton quantum beats is life-time of an individual exciton-polariton which is typically less than a hundred picoseconds even in high-quality semiconductor microcavities~\cite{PhysRevB.100.245304}. However, some oscillatory phenomena with much longer decay times in many-body bosonic condensates of exciton-polaritons have recently been theoretically discussed~\cite{PhysRevB.101.085302,PhysRevLett.114.193901,PhysRevA.99.033830} and experimentally observed~\cite{PhysRevLett.129.155301,barrat2023superfluid}.
\ak{Specifically, the persistent Larmor precession driven by polariton–polariton interactions has been reported~\cite{PhysRevLett.129.155301}, 
and the continuous time-crystal behavior has been demonstrated in a polariton condensate coupled to phonons~\cite{CarraroHaddad2024}.}
It has been argued that a combination of strong optical non-linearity caused by repulsive polariton-polariton interactions, stimulated scattering of polaritons and dissipative coupling between different polariton condensates may lead to the limit-cycle solutions~\cite{PhysRevB.101.085302}, dynamical attractors~\cite{PhysRevLett.114.193901,PhysRevB.109.155301} and even to polariton time-crystals~\cite{PhysRevA.99.033830}.  In contrast to the quantum beats characteristic of a linear two-level quantum system whose frequency is simply given by the splitting between two participating energy levels, the limit cycle oscillations have their dynamics governed by a variety of factors including the polariton concentration, polariton-polariton interaction constants, spatial dependence of the polariton lifetime $etc$~\cite{PhysRevLett.114.193901}. The decay time of oscillations characteristic of limit cycles, theoretically, tends to infinity.
In the present study, we experimentally observe long-lived oscillations of a macroscopic exciton-polariton condensate in real space that reveal features of quantum beats rather than limit-cycle oscillations.
\begin{figure*}
    \centering
    \includegraphics[width=0.6\linewidth]{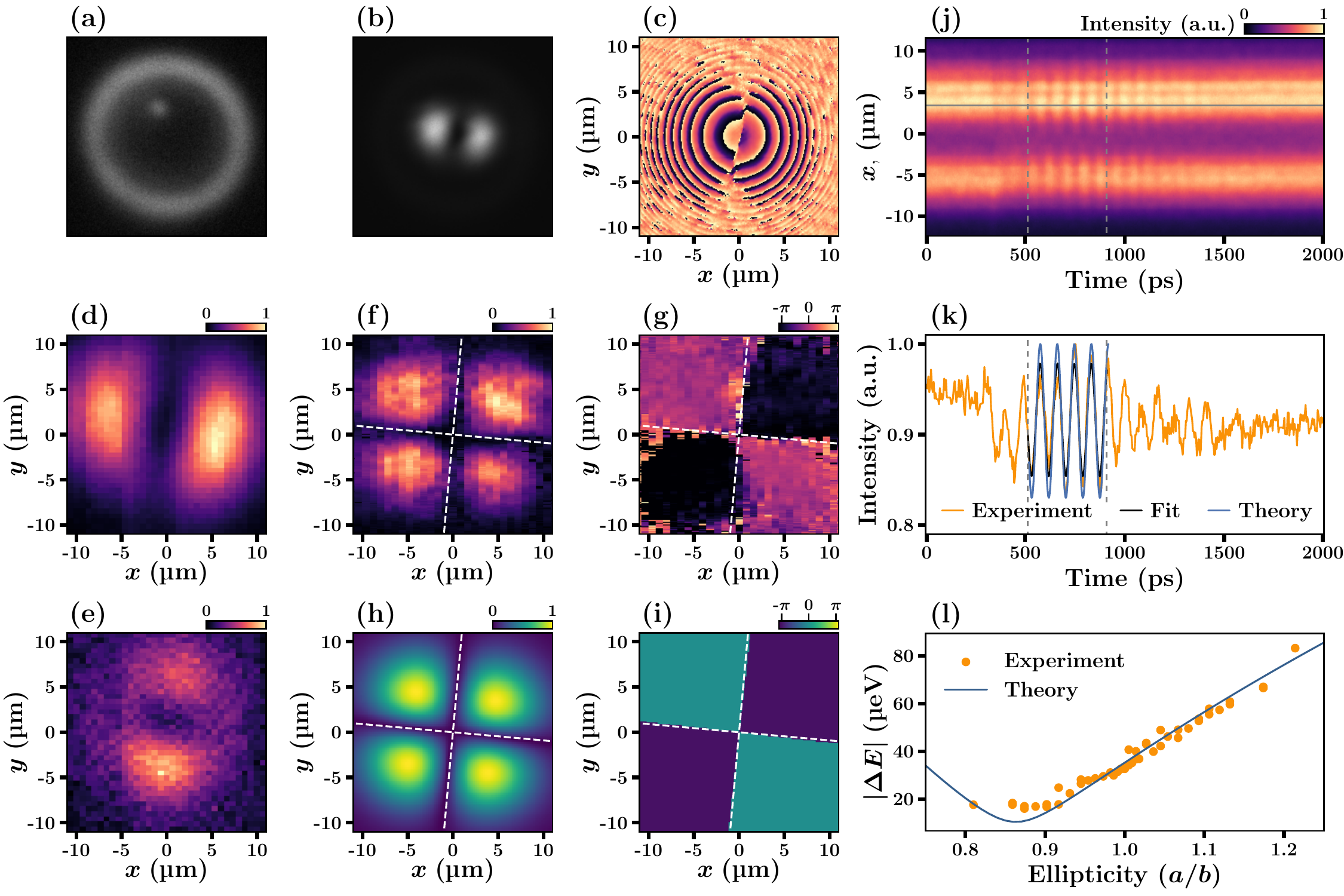}
    \caption{The optically-induced elliptical trap used for confinement of an exciton-polariton condensate. (a) the trap (light ring) and the spot of a control pulse inside the ring. (b) the intensity of emission of the trapped condensate in real space in the absence of quantum beats. (c) the space-dependent phase of the condensate wavefunction extracted from the interferometry images. Panels (d) and (e) show the experimental images of the eigen-functions of the trapped condensate referred to as $p_x$- and $p_y$- orbitals, corresponding to the 2nd and 3rd size-quantization energy levels of the trap. Panels (f), (g), (h), (i) show the maps of an amplitude $a(x,y)$ (f,h) and a phase $\phi_0(x,y)$ (g,i) of the oscillations obtained from fitting the experimental data (f,g) by harmonic oscillatory functions and calculated (h,i). Both in the experiment and theoretical calculation, the ellipticity coefficient of the trap of $1.054$ has been chosen. (j) shows the experimentally measured dynamics of the condensate emission intensity triggered by the control pulse and detected at the streak-camera slot oriented along $x$-axis. The orange line in (k) is a cut of the image in (j) along the horizontal line. The blue line in panel (k) shows the corresponding density of the condensate wavefunction calculated within the two-level model. (l) the energy splitting of $p_x$- and $p_y$-eigen energies $\delta E=\hbar\omega$ as a function of the ellipticity of the trap. Points show the splittings extracted from the fit of the experimental data. Blue line shows the calculation results. \ak{The fitting parameters of the model, entering the potential~\eqref{eqn:pot_basic}}: $m=7$ eV, $\omega_x=0.352$ ps$^{-1}$, $\omega_y=0.371$ ps$^{-1}$.}
    \label{fig:1}
\end{figure*}

\begin{figure*}
    \centering
    \includegraphics[width=0.6\linewidth]{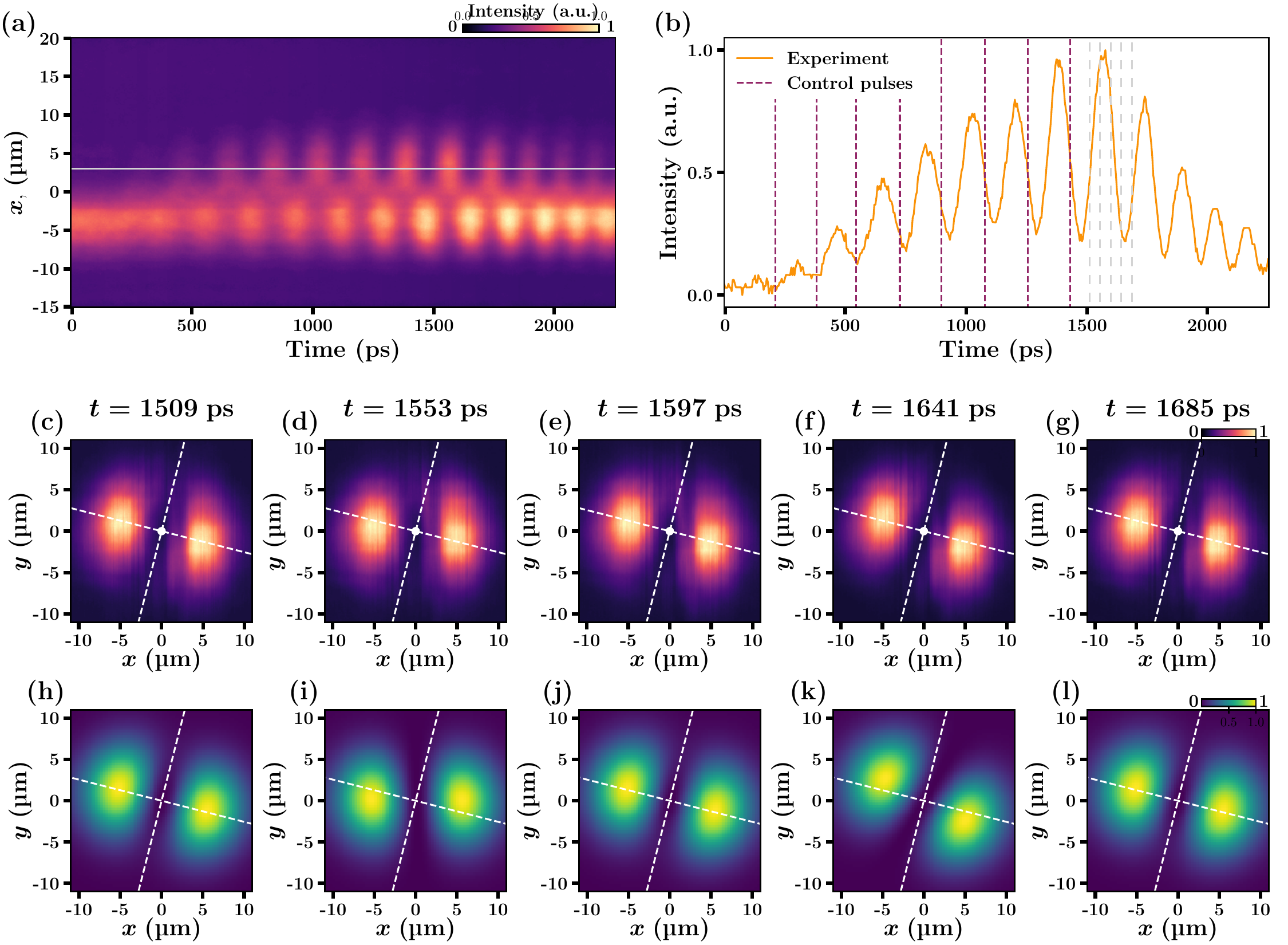}
    \caption{ (a) The time-dependent emission of the trapped polariton condensate detected within the streak-camera slit, (b) shows the cut of the image (a) along the horizontal white line indicated in (a). Vertical purple dashed lines show the times of arrival of the control pulses. Five dashed gray lines in (b) indicate the times at which the images of the polariton condensate in panels (c)--(g) are taken. The experimental images of the condensate emission intensity in real-space ((c)--(g)) demonstrate the evolution of the condensate wavefunction within a single period of oscillations, which is $176$ ps in this specific case. The corresponding numerically calculated images ((h)--(l)) are obtained with use of the set of the fitting parameters: $m=7$ eV, $\omega_x=0.297$ ps$^{-1}$, $\omega_y=0.303$ ps$^{-1}$.}
    \label{fig:2}
\end{figure*}

\textit{Setting the trap.---} We study bosonic condensates of exciton-polaritons created by a nonresonant $cw$ optical excitation in a planar GaAs-based microcavity. The description of the sample and its characteristic optical spectra can be found in ~\cite{Aladinskaia}. We use a spatial-light modulator (SLM) (see figure~\ref{Sfig:1} of the Supplementary Material) to create an elliptical trap for exciton-polaritons shown in figure~\ref{fig:1}, panel (a).
We take advantage of the driven-dissipative nature of exciton-polariton condensatesve bosonic condensation at energy states other than the ground state. We carefully chose the design of the trap and the pumping intensity in order to make sure that the ensemble of exciton-polaritons populates a selected pair of energy levels of the trapping potential. The eigen-wave functions corresponding to these states represent two-dimensional $p$-shaped orbitals. \ak{The corresponding spatial distributions of the intensity and the phase of the polariton condensate are shown on panels (b) and (c), respectively.}

To change the state of the condensate, or to put it into the oscillatory mode, we modify the profile of the potential trap with the use of nonresonant control pulses. These pulses excite electron and hole clouds that eventually form incoherent excitons that, in their turn, relax to the polariton modes. Together, this leads to the appearance of a localized repulsive potential acting on the polariton condensate. This potential eventually vanishes on a time scale of several hundred picoseconds (see figure~\ref{fig:7} of the Supplementary Material). The emission of excitons (exciton-polaritons) created by a control pulse can be directly observed with the use of the streak camera, as panel (a) in figure~\ref{fig:1} shows. One can see how its peak relaxes in energy, eventually reaching the condensate energy (figure~\ref{fig:7} of the Supplementary Material).

Elliptical traps offer the advantage of optical control over energy splitting between mentioned $p$-shaped orbitals. The experimentally obtained spatial distributions of the polariton densities corresponding to these orbitals are shown in panels (d) and (e) in figure~\ref{fig:1}. We map the true eigen functions of the trap by measuring the spatial maps of phase $\phi_0\left(x,y\right)$ and amplitude $a\left(x,y\right)$ of the oscillations of intensity of the emitted light induced by quantum beats between the eigen states of the trap. These maps obtained both experimentally and theoretically are shown in panels (f), (g), (h), and (i) of figure~\ref{fig:1}. The oscillations are the most pronounced in the areas of the strongest overlap of $p_x$ and $p_y$ orbitals. Varying the ellipticity of the trap we tune the splitting of $p_x$ and $p_y$ orbitals on a micro-electron-Volt scale as panel (l) in figure~\ref{fig:1} illustrates.

\textit{The dynamics of trapped condensates. ---} The oscillatory dynamics observed in our experiments can be analytically reproduced by a linear two-level model.  We approximate the optically induced trap that confines the exciton-polaritons condensate by a modified potential of a two-dimensional harmonic oscillator $V=m f(\varphi)\left(\omega_x^2 x^2+\omega_y^2 y^2\right)/2$ (for details, see the Supplementary Materials). We solve the two-dimensional Schr\"{o}dinger equation~\eqref{eqn:schrod_eq_sm} and obtain the energy spectrum of the trapped condensate. 
We assume that the system occupied the $p_x$-eigen state prior to the arrival of the control pulse. The effect of this non-resonant pulse we model by a time-dependent perturbation potential. Figure~\ref{fig:1}(k) shows the theoretical calculation of the density dynamics of the polariton condensate at the specific spot \ak{(horizontal cross-section in panel (j))} of the elliptical trap (blue curve). These results appear to be in good agreement with the experimental data (orange curve).

Streak-camera measurements allow us to visualize time- and space-resolved images of polariton condensates. We find a very good agreement of the measurements with predictions of a two-level quasi-analytical model. After reaching the stationary oscillation regime, we observe pronounced intensity beats which may be fitted by a harmonic function: $a(x,y) \sin{[\omega  t   + \phi_0(x,y)]}$ at each spatial point. Figure~\ref{fig:2} shows the spatial distribution of the polariton density obtained experimentally \ak{(panels (c)-(g))} and numerically \ak{(panels (h)-(l))} at different times within the period of quantum beats. In order to maximize the magnitude of intensity oscillations detected by streak-camera \ak{(panel (a))} we use a sequence of 8 control pulses incident to the same spot at equal time intervals, as dashed lines in figure~\ref{fig:2}(b) show. The repetition frequency in this sequence of pulses was chosen equal to the condensate oscillation frequency to achieve a parametric resonance (see Supplementary Material, figure~\ref{fig:9}). 

\begin{figure}
    \centering
    \includegraphics[width=1\linewidth]{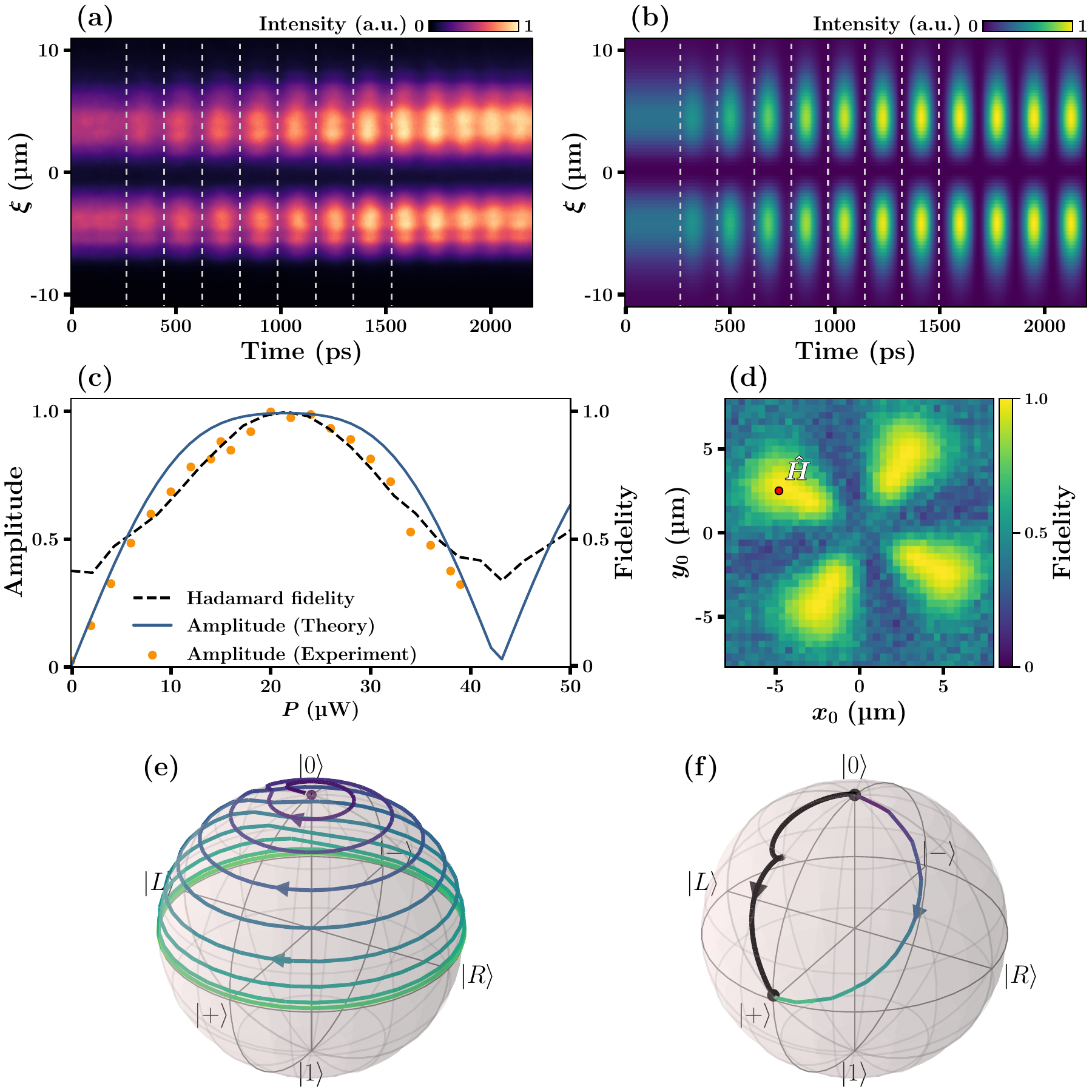}
    \caption{Implementation of the Hadamard operation~\eqref{eq:hadamar_gate_matrix} on a trapped polariton condensate. (a) the experimental streak-camera image of the condensate density cross-section as a function of time. Vertical dashed white lines show when the control laser pulses arrive. (b) the calculated dynamics of the condensate density cross-section. (c) the experimental (orange dots) and theoretical (blue solid line) dependencies of the amplitude of
the oscillations of the symmetry axis of the condensate on the pump power of control pulses. The black curve shows the dependence of the predicted fidelity of the Hadamard operation as a function of the control pulse intensity. (d) the color map showing the calculated dependence of the fidelity of the Hadamard gate on the coordinates $(x_0,y_0)$ of the spot hit by control pulses. The red dot shows the position used in the experiment. (e) shows the calculated trajectory on the surface of the Bloch sphere that describes the dynamics of the system initialized by 8 subsequent laser pulses, (f) shows the same trajectory in a frame rotating with the frequency of the observed quantum beats (blue-green line).\ak{The experimental dynamics of the system is shown by the black solid line in panel (f).} Parameters of the model: $m=7$ eV, $\omega_x=0.359$ ps$^{-1}$, $\omega_y=0.341$ ps$^{-1}$.}
    \label{fig:3}
\end{figure}
\textit{Hadamard and Pauli operations.---} 
It is instructive to map the observed quantum beats to a Bloch sphere. We do it by fitting the time-resolved tomography images recorded by the streak-camera with use of a two-level model introduced above. The oscillatory dynamics of the system manifests itself in the precession of its quantum state on the surface of the Bloch sphere as panel \ak{(e) (in a rest frame) and (f) (in a rotating frame)} in figure~\ref{fig:3} show. 

The poles of the Bloch sphere may be considered as the computational basis states $|0\rangle$ \ak{($p_x$-orbital)} and $|1\rangle$ \ak{($p_y$-orbital)}. Once a polariton condensate is placed in a superposition of two eigen states of a trap, it may be considered as a qubit. Using nonresonant optical pulses focused at different locations of the trap, we were able to change the state of the trapped condensate. These changes manifest themselves in changing frequency, amplitude and phase of the quantum beats observed in the streak-camera images. Mapping the state of the condensate to the Bloch sphere we observed how the studied quantum system evolves from the pole of the sphere to its equatorial plane (figure~\ref{fig:3}(e,f)). \ak{Panel (f) displays two curves: the experimental data is represented by the black curve, whereas the theoretical calculation is shown in blue-green. The two trajectories converge to the same final point, which is crucial for realizing a high fidelity quantum operation.} This particular transformation corresponds to the Hadamard operation given by:
\begin{align}\label{eq:hadamar_gate_matrix}
H = \frac{1}{\sqrt{2}} \begin{bmatrix} 1 & 1 \\ 1 & -1 \end{bmatrix}.
\end{align}
%
In order to implement the high-fidelity Hadamard operation, we have measured the dependence of the amplitude of the beats induced by the sequence of control pulses on the pump power and compared it with the theoretical prediction (figure~\ref{fig:3}(c)). The black line in figure~\ref{fig:3}(c) shows the predicted fidelity of the Hadamard operation for a pulse centered at the red spot in the panel (d). 
Carefully choosing the spot hit by the sequence of control pulses we have been able to achieve a fidelity of the Hadamard operation of over $0.95$. \ak{This number is obtained by averaging over 5 different initial states of the system}. The theoretical estimate for the fidelity is much higher, about $\sim 1$, as one can see from the calculated fidelity map shown in figure~\ref{fig:3}(d). 

\begin{figure}
    \centering
    \includegraphics[width=1\linewidth]{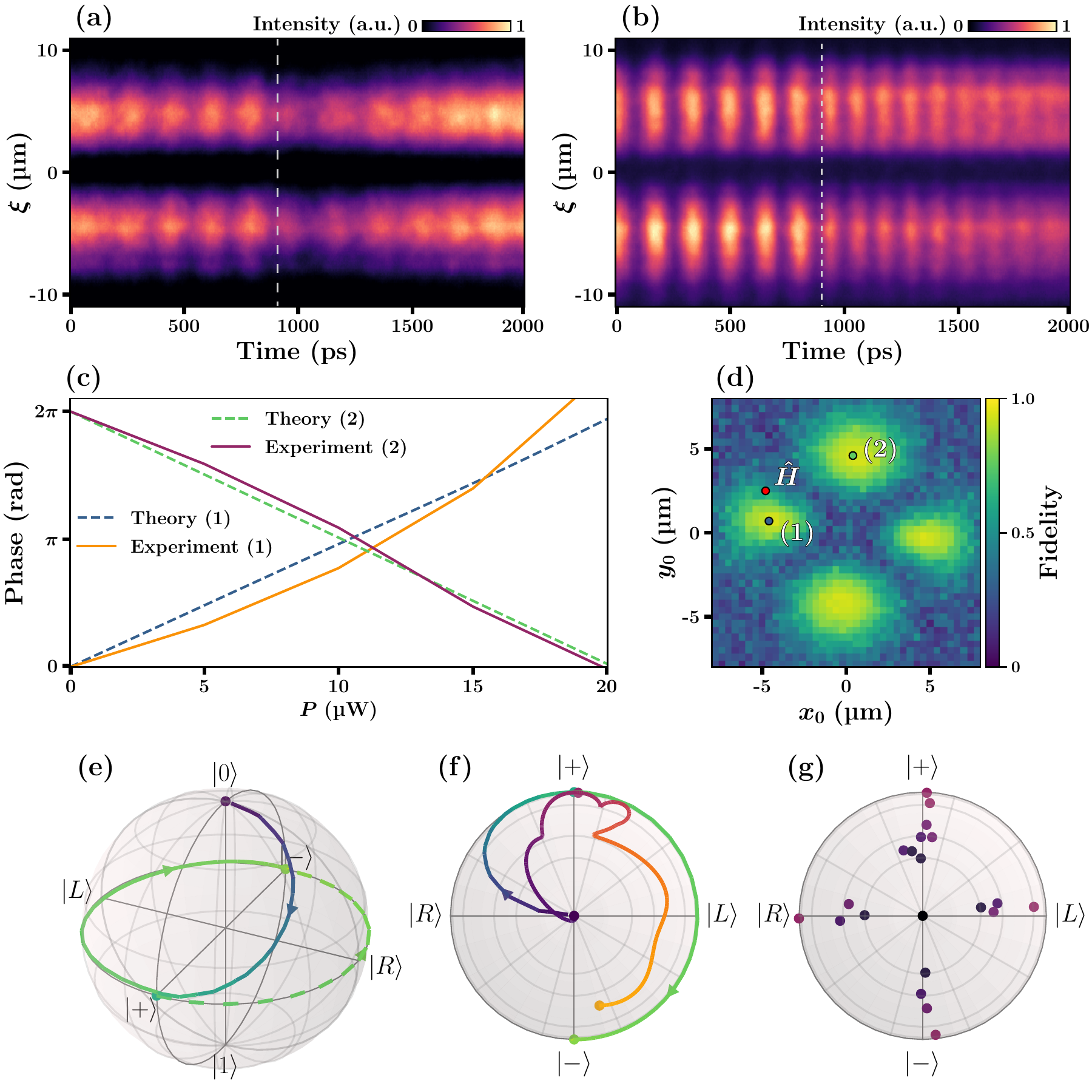}
    \caption{Implementation of the Pauli-$Z$ operation~\eqref{eq:pauliz_gate_matrix} on a trapped polariton condensate. (a) the streak-camera image of the condensate density cross-section as a function of time for the control pulse location shown by dot (1) in panel (d), (b) shows the same for the control pulse location shown by dot (2) in panel (d). The arrival times of control pulses are indicated by vertical dashed lines in panels (a) and (b). (c) the experimental (orange and purple lines) and  theoretical (dashed blue and green lines) dependencies of the shift of the phase of quantum beats induced by control pulses incident to the sample at the locations (1) and (2) indicated in panel (d) as functions of the corresponding pumping intensities. (d) color map showing the calculated dependence of the fidelity of the Pauli-$Z$ gate on the coordinates $(x_0,y_0)$ of the spot hit by control pulses. The red spot shows location of the first 8 pulses used to implement the Hadamard operation, the blue and green spots marked (1) and (2) show locations of the 9th pulse triggering the Pauli-$Z$ operation. \ak{(e) shows the calculated trajectories of the system on the surface of the Bloch sphere for the combination of 8 pulses and 9th pulse arriving at locations (1) and (2) shown in the panel (d) by solid and dotted lines, respectively. (f) demonstrates the  experimental dynamics of the system (orange-purple line) subject to the   Hadamard and Pauli-$Z$ gate operations. The theoretical trajectory from (e) is also shown for comparison. (g) shows the set of examined initial states of the system used to evaluate fidelity of the Pauli-$Z$ operation. Parameters of the model: $m=7$ eV, $\omega_x=0.152$ ps$^{-1}$, $\omega_y=0.164$ ps$^{-1}$.}} 
    \label{fig:4}
\end{figure}

Once the Hadamard operation is implemented, we implement also Pauli-$Z$ operation given by:
\begin{align}\label{eq:pauliz_gate_matrix}
Z = \begin{bmatrix} 1 & 0 \\ 0 & -1 \end{bmatrix}.
\end{align}
\ak{This is achieved by sending the 9th control pulse to a specific spatial location. It moves the system to an opposite end of the diameter in the equatorial plane of the Bloch sphere, as figure~\ref{fig:4} illustrates. The streak-camera images of the condensate density cross-section as a function of time are demonstrated in panels (a) and (b). 8 initializing pulses transfer the system from the pole to the equator ($\ket{+}$), then the 9th control pulse triggers the Pauli-$Z$ gate operation. We have plotted the dependencies of the phase shifts induced by the control pulses as functions of the control pulse intensities (panel (c)). After optimization of the pump power the phase of the quantum beats is being shifted by this gate operation by $+\pi$ or $-\pi$ depending on the location of the control pulse. The final state of the system is the same in both cases, as panel (e) illustrates. This panel shows two calculated trajectories of the system, which coincide on the path from $\ket{0}$ to $\ket{+}$, and then move in opposite directions along the equator (solid and dotted lines). The experimental fidelity of the implemented Pauli-Z operation exceeds $0.97$ in both cases. We have obtained this estimate based on a series of experiments, where we initialized the system in $>20$ different points on the surface of the Bloch sphere. The studied set of initial locations is shown in panel (g). }

\textit{Discussion.---} In our experiments, streak-camera averages the emission intensity collected after millions of pulses. Still, we are able to observe the intensity beats whose amplitude and phase are characterized by well-defined patterns.  The macroscopic many-body wavefunction of a polariton condensate changes on a length-scale of over $\SI{10}{\micro\metre}$ with a periodicity of about $100~\textup{ps}$.
This indicates that every initial pulse brings the condensate essentially to the same superposition of $p_x$- and $p_y$-orbitals. Regular oscillations of a trapped exciton-polariton condensate that persist for at least $1~\textup{ns}$. The phase locking of $p_x$- and $p_y$-orbitals imposed by the control pulse survives much longer than the phase of the condensate as a whole. This conclusion is consistent with recent experiments on persistent currents of exciton-polaritons in a ring geometry ~\cite{Snoke}. \ak{We note also that the real-space density dynamics of polariton condensates in traps have been studied in coupled systems~\cite{PhysRevB.102.195428,PhysRevLett.126.075301}. These studies focused on nonlinear phenomena such as Lotka-Volterra population dynamics or collective Bogoliubov-like modes, where the interplay of interactions and dissipation drives pronounced density oscillations. In contrast, our present experiment demonstrates a linear phenomenon of quantum beats between the discrete $p_x$ and $p_y$ orbitals.}  \ak{Barrat \textit{et al.}~\cite{barrat2023superfluid} have recently implemented a polariton-based qubit analog in an annular trap, exploiting two counter-circulating vortex states. Further exploring the potentiality of trapped condensates for quantum computation, here we harness the linear splitting between $p_x$ and $p_y$ orbitals confined by an elliptical trap and directly visualize the dynamics of the condensate wave function in real space. The frequency of the oscillations that we observe is defined by the splitting between size-quantization levels of the polariton condensate confined in a trap and it can be tuned either by changing the ellipticity of the trap or by modifying its potential with use of control pulses of light. In order to check that the observed dynamics is not linked with any dynamical attractor of a limit-cycle type, we excited the system in several initial states on the Bloch sphere. Shifting the initial condition we shift the whole trajectory on a sphere. Furthermore, we have perturbed the dynamics of the condensate by sending control laser pulses. In these studies, no trace of an attractor has been revealed.} We conclude that the observed oscillations are quantum beats experienced by the polariton condensate as a whole entity. These beats have several unique features: (i) they persist over times orders of magnitude longer than the single-polariton lifetime, (ii) they involve several thousands of polaritons composing the condensate, (iii) a pronounced real space dynamics is observed.  The observation of coherent real-space dynamics of a macroscopic many-body object paves the way for the realization of polariton qubits, as discussed in recent publications~\cite{barrat2023superfluid,PhysRevResearch.3.013099,Kavokin2022,Ricco2024}.

\textit{Acknowledgements.---} The authors acknowledge the Saint Petersburg State University for the Research Grant No. 125022803069-4. AVK acknowledges "Innovation Program for Quantum Science and Technology" 2023ZD0300300. RC acknowledges the Resource Center Nanophotonics of the State University of St-Petersburg for support of the experimental part of the project. AK acknowledges support from the Moscow Institute of Physics and Technology under the Priority 2030 Strategic Academic Leadership Program.

\bibliography{references}

\begin{thebibliography}{47}%
\makeatletter
\providecommand \@ifxundefined [1]{%
 \@ifx{#1\undefined}
}%
\providecommand \@ifnum [1]{%
 \ifnum #1\expandafter \@firstoftwo
 \else \expandafter \@secondoftwo
 \fi
}%
\providecommand \@ifx [1]{%
 \ifx #1\expandafter \@firstoftwo
 \else \expandafter \@secondoftwo
 \fi
}%
\providecommand \natexlab [1]{#1}%
\providecommand \enquote  [1]{``#1''}%
\providecommand \bibnamefont  [1]{#1}%
\providecommand \bibfnamefont [1]{#1}%
\providecommand \citenamefont [1]{#1}%
\providecommand \href@noop [0]{\@secondoftwo}%
\providecommand \href [0]{\begingroup \@sanitize@url \@href}%
\providecommand \@href[1]{\@@startlink{#1}\@@href}%
\providecommand \@@href[1]{\endgroup#1\@@endlink}%
\providecommand \@sanitize@url [0]{\catcode `\\12\catcode `\$12\catcode `\&12\catcode `\#12\catcode `\^12\catcode `\_12\catcode `\%12\relax}%
\providecommand \@@startlink[1]{}%
\providecommand \@@endlink[0]{}%
\providecommand \url  [0]{\begingroup\@sanitize@url \@url }%
\providecommand \@url [1]{\endgroup\@href {#1}{\urlprefix }}%
\providecommand \urlprefix  [0]{URL }%
\providecommand \Eprint [0]{\href }%
\providecommand \doibase [0]{https://doi.org/}%
\providecommand \selectlanguage [0]{\@gobble}%
\providecommand \bibinfo  [0]{\@secondoftwo}%
\providecommand \bibfield  [0]{\@secondoftwo}%
\providecommand \translation [1]{[#1]}%
\providecommand \BibitemOpen [0]{}%
\providecommand \bibitemStop [0]{}%
\providecommand \bibitemNoStop [0]{.\EOS\space}%
\providecommand \EOS [0]{\spacefactor3000\relax}%
\providecommand \BibitemShut  [1]{\csname bibitem#1\endcsname}%
\let\auto@bib@innerbib\@empty
\bibitem [{\citenamefont {Mitsunaga}\ and\ \citenamefont {Tang}(1987)}]{PhysRevA.35.1720}%
  \BibitemOpen
  \bibfield  {author} {\bibinfo {author} {\bibfnamefont {M.}~\bibnamefont {Mitsunaga}}\ and\ \bibinfo {author} {\bibfnamefont {C.~L.}\ \bibnamefont {Tang}},\ }\href {https://doi.org/10.1103/PhysRevA.35.1720} {\bibfield  {journal} {\bibinfo  {journal} {Phys. Rev. A}\ }\textbf {\bibinfo {volume} {35}},\ \bibinfo {pages} {1720} (\bibinfo {year} {1987})}\BibitemShut {NoStop}%
\bibitem [{\citenamefont {Dodd}\ \emph {et~al.}(1964)\citenamefont {Dodd}, \citenamefont {Kaul},\ and\ \citenamefont {Warrington}}]{J_N_Dodd_1964}%
  \BibitemOpen
  \bibfield  {author} {\bibinfo {author} {\bibfnamefont {J.~N.}\ \bibnamefont {Dodd}}, \bibinfo {author} {\bibfnamefont {R.~D.}\ \bibnamefont {Kaul}},\ and\ \bibinfo {author} {\bibfnamefont {D.~M.}\ \bibnamefont {Warrington}},\ }\href {https://doi.org/10.1088/0370-1328/84/1/123} {\bibfield  {journal} {\bibinfo  {journal} {Proceedings of the Physical Society}\ }\textbf {\bibinfo {volume} {84}},\ \bibinfo {pages} {176} (\bibinfo {year} {1964})}\BibitemShut {NoStop}%
\bibitem [{\citenamefont {Haroche}\ \emph {et~al.}(1973)\citenamefont {Haroche}, \citenamefont {Paisner},\ and\ \citenamefont {Schawlow}}]{PhysRevLett.30.948}%
  \BibitemOpen
  \bibfield  {author} {\bibinfo {author} {\bibfnamefont {S.}~\bibnamefont {Haroche}}, \bibinfo {author} {\bibfnamefont {J.~A.}\ \bibnamefont {Paisner}},\ and\ \bibinfo {author} {\bibfnamefont {A.~L.}\ \bibnamefont {Schawlow}},\ }\href {https://doi.org/10.1103/PhysRevLett.30.948} {\bibfield  {journal} {\bibinfo  {journal} {Phys. Rev. Lett.}\ }\textbf {\bibinfo {volume} {30}},\ \bibinfo {pages} {948} (\bibinfo {year} {1973})}\BibitemShut {NoStop}%
\bibitem [{\citenamefont {Shah}(2013)}]{shah2013ultrafast}%
  \BibitemOpen
  \bibfield  {author} {\bibinfo {author} {\bibfnamefont {J.}~\bibnamefont {Shah}},\ }\href {https://doi.org/10.1007/978-3-662-03770-6} {\emph {\bibinfo {title} {Ultrafast spectroscopy of semiconductors and semiconductor nanostructures, 2nd ed.}}},\ Vol.\ \bibinfo {volume} {115}\ (\bibinfo  {publisher} {Springer},\ \bibinfo {address} {New York, USA},\ \bibinfo {year} {2013})\BibitemShut {NoStop}%
\bibitem [{\citenamefont {Roskos}\ \emph {et~al.}(1992)\citenamefont {Roskos}, \citenamefont {Nuss}, \citenamefont {Shah}, \citenamefont {Leo}, \citenamefont {Miller}, \citenamefont {Fox}, \citenamefont {Schmitt-Rink},\ and\ \citenamefont {K\"ohler}}]{PhysRevLett.68.2216}%
  \BibitemOpen
  \bibfield  {author} {\bibinfo {author} {\bibfnamefont {H.~G.}\ \bibnamefont {Roskos}}, \bibinfo {author} {\bibfnamefont {M.~C.}\ \bibnamefont {Nuss}}, \bibinfo {author} {\bibfnamefont {J.}~\bibnamefont {Shah}}, \bibinfo {author} {\bibfnamefont {K.}~\bibnamefont {Leo}}, \bibinfo {author} {\bibfnamefont {D.~A.~B.}\ \bibnamefont {Miller}}, \bibinfo {author} {\bibfnamefont {A.~M.}\ \bibnamefont {Fox}}, \bibinfo {author} {\bibfnamefont {S.}~\bibnamefont {Schmitt-Rink}},\ and\ \bibinfo {author} {\bibfnamefont {K.}~\bibnamefont {K\"ohler}},\ }\href {https://doi.org/10.1103/PhysRevLett.68.2216} {\bibfield  {journal} {\bibinfo  {journal} {Phys. Rev. Lett.}\ }\textbf {\bibinfo {volume} {68}},\ \bibinfo {pages} {2216} (\bibinfo {year} {1992})}\BibitemShut {NoStop}%
\bibitem [{\citenamefont {Schmitt-Rink}\ \emph {et~al.}(1992)\citenamefont {Schmitt-Rink}, \citenamefont {Bennhardt}, \citenamefont {Heuckeroth}, \citenamefont {Thomas}, \citenamefont {Haring}, \citenamefont {Maidorn}, \citenamefont {Bakker}, \citenamefont {Leo}, \citenamefont {Kim}, \citenamefont {Shah},\ and\ \citenamefont {K\"ohler}}]{PhysRevB.46.10460}%
  \BibitemOpen
  \bibfield  {author} {\bibinfo {author} {\bibfnamefont {S.}~\bibnamefont {Schmitt-Rink}}, \bibinfo {author} {\bibfnamefont {D.}~\bibnamefont {Bennhardt}}, \bibinfo {author} {\bibfnamefont {V.}~\bibnamefont {Heuckeroth}}, \bibinfo {author} {\bibfnamefont {P.}~\bibnamefont {Thomas}}, \bibinfo {author} {\bibfnamefont {P.}~\bibnamefont {Haring}}, \bibinfo {author} {\bibfnamefont {G.}~\bibnamefont {Maidorn}}, \bibinfo {author} {\bibfnamefont {H.}~\bibnamefont {Bakker}}, \bibinfo {author} {\bibfnamefont {K.}~\bibnamefont {Leo}}, \bibinfo {author} {\bibfnamefont {D.-S.}\ \bibnamefont {Kim}}, \bibinfo {author} {\bibfnamefont {J.}~\bibnamefont {Shah}},\ and\ \bibinfo {author} {\bibfnamefont {K.}~\bibnamefont {K\"ohler}},\ }\href {https://doi.org/10.1103/PhysRevB.46.10460} {\bibfield  {journal} {\bibinfo  {journal} {Phys. Rev. B}\ }\textbf {\bibinfo {volume} {46}},\ \bibinfo {pages} {10460} (\bibinfo {year} {1992})}\BibitemShut {NoStop}%
\bibitem [{\citenamefont {G\"obel}\ \emph {et~al.}(1990)\citenamefont {G\"obel}, \citenamefont {Leo}, \citenamefont {Damen}, \citenamefont {Shah}, \citenamefont {Schmitt-Rink}, \citenamefont {Sch\"afer}, \citenamefont {M\"uller},\ and\ \citenamefont {K\"ohler}}]{PhysRevLett.64.1801}%
  \BibitemOpen
  \bibfield  {author} {\bibinfo {author} {\bibfnamefont {E.~O.}\ \bibnamefont {G\"obel}}, \bibinfo {author} {\bibfnamefont {K.}~\bibnamefont {Leo}}, \bibinfo {author} {\bibfnamefont {T.~C.}\ \bibnamefont {Damen}}, \bibinfo {author} {\bibfnamefont {J.}~\bibnamefont {Shah}}, \bibinfo {author} {\bibfnamefont {S.}~\bibnamefont {Schmitt-Rink}}, \bibinfo {author} {\bibfnamefont {W.}~\bibnamefont {Sch\"afer}}, \bibinfo {author} {\bibfnamefont {J.~F.}\ \bibnamefont {M\"uller}},\ and\ \bibinfo {author} {\bibfnamefont {K.}~\bibnamefont {K\"ohler}},\ }\href {https://doi.org/10.1103/PhysRevLett.64.1801} {\bibfield  {journal} {\bibinfo  {journal} {Phys. Rev. Lett.}\ }\textbf {\bibinfo {volume} {64}},\ \bibinfo {pages} {1801} (\bibinfo {year} {1990})}\BibitemShut {NoStop}%
\bibitem [{\citenamefont {Bar-Ad}\ and\ \citenamefont {Bar-Joseph}(1991)}]{PhysRevLett.66.2491}%
  \BibitemOpen
  \bibfield  {author} {\bibinfo {author} {\bibfnamefont {S.}~\bibnamefont {Bar-Ad}}\ and\ \bibinfo {author} {\bibfnamefont {I.}~\bibnamefont {Bar-Joseph}},\ }\href {https://doi.org/10.1103/PhysRevLett.66.2491} {\bibfield  {journal} {\bibinfo  {journal} {Phys. Rev. Lett.}\ }\textbf {\bibinfo {volume} {66}},\ \bibinfo {pages} {2491} (\bibinfo {year} {1991})}\BibitemShut {NoStop}%
\bibitem [{\citenamefont {Trifonov}\ \emph {et~al.}(2019)\citenamefont {Trifonov}, \citenamefont {Khramtsov}, \citenamefont {Kavokin}, \citenamefont {Ignatiev}, \citenamefont {Kavokin}, \citenamefont {Efimov}, \citenamefont {Eliseev}, \citenamefont {Shapochkin},\ and\ \citenamefont {Bayer}}]{PhysRevLett.122.147401}%
  \BibitemOpen
  \bibfield  {author} {\bibinfo {author} {\bibfnamefont {A.~V.}\ \bibnamefont {Trifonov}}, \bibinfo {author} {\bibfnamefont {E.~S.}\ \bibnamefont {Khramtsov}}, \bibinfo {author} {\bibfnamefont {K.~V.}\ \bibnamefont {Kavokin}}, \bibinfo {author} {\bibfnamefont {I.~V.}\ \bibnamefont {Ignatiev}}, \bibinfo {author} {\bibfnamefont {A.~V.}\ \bibnamefont {Kavokin}}, \bibinfo {author} {\bibfnamefont {Y.~P.}\ \bibnamefont {Efimov}}, \bibinfo {author} {\bibfnamefont {S.~A.}\ \bibnamefont {Eliseev}}, \bibinfo {author} {\bibfnamefont {P.~Y.}\ \bibnamefont {Shapochkin}},\ and\ \bibinfo {author} {\bibfnamefont {M.}~\bibnamefont {Bayer}},\ }\href {https://doi.org/10.1103/PhysRevLett.122.147401} {\bibfield  {journal} {\bibinfo  {journal} {Phys. Rev. Lett.}\ }\textbf {\bibinfo {volume} {122}},\ \bibinfo {pages} {147401} (\bibinfo {year} {2019})}\BibitemShut {NoStop}%
\bibitem [{\citenamefont {Kotur}\ \emph {et~al.}(2016)\citenamefont {Kotur}, \citenamefont {Dzhioev}, \citenamefont {Vladimirova}, \citenamefont {Jouault}, \citenamefont {Korenev},\ and\ \citenamefont {Kavokin}}]{PhysRevB.94.081201}%
  \BibitemOpen
  \bibfield  {author} {\bibinfo {author} {\bibfnamefont {M.}~\bibnamefont {Kotur}}, \bibinfo {author} {\bibfnamefont {R.~I.}\ \bibnamefont {Dzhioev}}, \bibinfo {author} {\bibfnamefont {M.}~\bibnamefont {Vladimirova}}, \bibinfo {author} {\bibfnamefont {B.}~\bibnamefont {Jouault}}, \bibinfo {author} {\bibfnamefont {V.~L.}\ \bibnamefont {Korenev}},\ and\ \bibinfo {author} {\bibfnamefont {K.~V.}\ \bibnamefont {Kavokin}},\ }\href {https://doi.org/10.1103/PhysRevB.94.081201} {\bibfield  {journal} {\bibinfo  {journal} {Phys. Rev. B}\ }\textbf {\bibinfo {volume} {94}},\ \bibinfo {pages} {081201} (\bibinfo {year} {2016})}\BibitemShut {NoStop}%
\bibitem [{\citenamefont {Alexandrov}(1964)}]{alexandrov1964interference}%
  \BibitemOpen
  \bibfield  {author} {\bibinfo {author} {\bibfnamefont {E.}~\bibnamefont {Alexandrov}},\ }\href@noop {} {\bibfield  {journal} {\bibinfo  {journal} {Opt. Spectrosc}\ }\textbf {\bibinfo {volume} {17}},\ \bibinfo {pages} {957} (\bibinfo {year} {1964})}\BibitemShut {NoStop}%
\bibitem [{\citenamefont {Erland}\ and\ \citenamefont {Balslev}(1993)}]{PhysRevA.48.R1765}%
  \BibitemOpen
  \bibfield  {author} {\bibinfo {author} {\bibfnamefont {J.}~\bibnamefont {Erland}}\ and\ \bibinfo {author} {\bibfnamefont {I.}~\bibnamefont {Balslev}},\ }\href {https://doi.org/10.1103/PhysRevA.48.R1765} {\bibfield  {journal} {\bibinfo  {journal} {Phys. Rev. A}\ }\textbf {\bibinfo {volume} {48}},\ \bibinfo {pages} {R1765} (\bibinfo {year} {1993})}\BibitemShut {NoStop}%
\bibitem [{\citenamefont {Yugova}\ \emph {et~al.}(2009)\citenamefont {Yugova}, \citenamefont {Glazov}, \citenamefont {Ivchenko},\ and\ \citenamefont {Efros}}]{PhysRevB.80.104436}%
  \BibitemOpen
  \bibfield  {author} {\bibinfo {author} {\bibfnamefont {I.~A.}\ \bibnamefont {Yugova}}, \bibinfo {author} {\bibfnamefont {M.~M.}\ \bibnamefont {Glazov}}, \bibinfo {author} {\bibfnamefont {E.~L.}\ \bibnamefont {Ivchenko}},\ and\ \bibinfo {author} {\bibfnamefont {A.~L.}\ \bibnamefont {Efros}},\ }\href {https://doi.org/10.1103/PhysRevB.80.104436} {\bibfield  {journal} {\bibinfo  {journal} {Phys. Rev. B}\ }\textbf {\bibinfo {volume} {80}},\ \bibinfo {pages} {104436} (\bibinfo {year} {2009})}\BibitemShut {NoStop}%
\bibitem [{\citenamefont {M\"uller}\ \emph {et~al.}(2010)\citenamefont {M\"uller}, \citenamefont {Oestreich}, \citenamefont {R\"omer},\ and\ \citenamefont {H\"ubner}}]{MULLER2010569}%
  \BibitemOpen
  \bibfield  {author} {\bibinfo {author} {\bibfnamefont {G.~M.}\ \bibnamefont {M\"uller}}, \bibinfo {author} {\bibfnamefont {M.}~\bibnamefont {Oestreich}}, \bibinfo {author} {\bibfnamefont {M.}~\bibnamefont {R\"omer}},\ and\ \bibinfo {author} {\bibfnamefont {J.}~\bibnamefont {H\"ubner}},\ }\href {https://doi.org/https://doi.org/10.1016/j.physe.2010.08.010} {\bibfield  {journal} {\bibinfo  {journal} {Physica E: Low-dimensional Systems and Nanostructures}\ }\textbf {\bibinfo {volume} {43}},\ \bibinfo {pages} {569} (\bibinfo {year} {2010})}\BibitemShut {NoStop}%
\bibitem [{\citenamefont {Kurnit}\ \emph {et~al.}(1964)\citenamefont {Kurnit}, \citenamefont {Abella},\ and\ \citenamefont {Hartmann}}]{PhysRevLett.13.567}%
  \BibitemOpen
  \bibfield  {author} {\bibinfo {author} {\bibfnamefont {N.~A.}\ \bibnamefont {Kurnit}}, \bibinfo {author} {\bibfnamefont {I.~D.}\ \bibnamefont {Abella}},\ and\ \bibinfo {author} {\bibfnamefont {S.~R.}\ \bibnamefont {Hartmann}},\ }\href {https://doi.org/10.1103/PhysRevLett.13.567} {\bibfield  {journal} {\bibinfo  {journal} {Phys. Rev. Lett.}\ }\textbf {\bibinfo {volume} {13}},\ \bibinfo {pages} {567} (\bibinfo {year} {1964})}\BibitemShut {NoStop}%
\bibitem [{\citenamefont {Scully}\ and\ \citenamefont {Zubairy}(1997)}]{Scully1997-aq}%
  \BibitemOpen
  \bibfield  {author} {\bibinfo {author} {\bibfnamefont {M.~O.}\ \bibnamefont {Scully}}\ and\ \bibinfo {author} {\bibfnamefont {M.~S.}\ \bibnamefont {Zubairy}},\ }in\ \href@noop {} {\emph {\bibinfo {booktitle} {Quantum Optics}}}\ (\bibinfo  {publisher} {Cambridge University Press},\ \bibinfo {address} {Cambridge},\ \bibinfo {year} {1997})\ pp.\ \bibinfo {pages} {17--18}\BibitemShut {NoStop}%
\bibitem [{\citenamefont {Deng}\ \emph {et~al.}(2002)\citenamefont {Deng}, \citenamefont {Weihs}, \citenamefont {Santori}, \citenamefont {Bloch},\ and\ \citenamefont {Yamamoto}}]{Deng2002}%
  \BibitemOpen
  \bibfield  {author} {\bibinfo {author} {\bibfnamefont {H.}~\bibnamefont {Deng}}, \bibinfo {author} {\bibfnamefont {G.}~\bibnamefont {Weihs}}, \bibinfo {author} {\bibfnamefont {C.}~\bibnamefont {Santori}}, \bibinfo {author} {\bibfnamefont {J.}~\bibnamefont {Bloch}},\ and\ \bibinfo {author} {\bibfnamefont {Y.}~\bibnamefont {Yamamoto}},\ }\href {https://doi.org/10.1126/science.1074464} {\bibfield  {journal} {\bibinfo  {journal} {Science}\ }\textbf {\bibinfo {volume} {298}},\ \bibinfo {pages} {199–202} (\bibinfo {year} {2002})}\BibitemShut {NoStop}%
\bibitem [{\citenamefont {Kasprzak}\ \emph {et~al.}(2006)\citenamefont {Kasprzak}, \citenamefont {Richard}, \citenamefont {Kundermann}, \citenamefont {Baas}, \citenamefont {Jeambrun}, \citenamefont {Keeling}, \citenamefont {Marchetti}, \citenamefont {Szymańska}, \citenamefont {André}, \citenamefont {Staehli}, \citenamefont {Savona}, \citenamefont {Littlewood}, \citenamefont {Deveaud},\ and\ \citenamefont {Dang}}]{Kasprzak2006}%
  \BibitemOpen
  \bibfield  {author} {\bibinfo {author} {\bibfnamefont {J.}~\bibnamefont {Kasprzak}}, \bibinfo {author} {\bibfnamefont {M.}~\bibnamefont {Richard}}, \bibinfo {author} {\bibfnamefont {S.}~\bibnamefont {Kundermann}}, \bibinfo {author} {\bibfnamefont {A.}~\bibnamefont {Baas}}, \bibinfo {author} {\bibfnamefont {P.}~\bibnamefont {Jeambrun}}, \bibinfo {author} {\bibfnamefont {J.~M.~J.}\ \bibnamefont {Keeling}}, \bibinfo {author} {\bibfnamefont {F.~M.}\ \bibnamefont {Marchetti}}, \bibinfo {author} {\bibfnamefont {M.~H.}\ \bibnamefont {Szymańska}}, \bibinfo {author} {\bibfnamefont {R.}~\bibnamefont {André}}, \bibinfo {author} {\bibfnamefont {J.~L.}\ \bibnamefont {Staehli}}, \bibinfo {author} {\bibfnamefont {V.}~\bibnamefont {Savona}}, \bibinfo {author} {\bibfnamefont {P.~B.}\ \bibnamefont {Littlewood}}, \bibinfo {author} {\bibfnamefont {B.}~\bibnamefont {Deveaud}},\ and\ \bibinfo {author} {\bibfnamefont {L.~S.}\ \bibnamefont {Dang}},\ }\href {https://doi.org/10.1038/nature05131} {\bibfield  {journal} {\bibinfo
  {journal} {Nature}\ }\textbf {\bibinfo {volume} {443}},\ \bibinfo {pages} {409–414} (\bibinfo {year} {2006})}\BibitemShut {NoStop}%
\bibitem [{\citenamefont {Carusotto}\ and\ \citenamefont {Ciuti}(2013)}]{RevModPhys.85.299}%
  \BibitemOpen
  \bibfield  {author} {\bibinfo {author} {\bibfnamefont {I.}~\bibnamefont {Carusotto}}\ and\ \bibinfo {author} {\bibfnamefont {C.}~\bibnamefont {Ciuti}},\ }\href {https://doi.org/10.1103/RevModPhys.85.299} {\bibfield  {journal} {\bibinfo  {journal} {Rev. Mod. Phys.}\ }\textbf {\bibinfo {volume} {85}},\ \bibinfo {pages} {299} (\bibinfo {year} {2013})}\BibitemShut {NoStop}%
\bibitem [{\citenamefont {Norris}\ \emph {et~al.}(1994)\citenamefont {Norris}, \citenamefont {Rhee}, \citenamefont {Sung}, \citenamefont {Arakawa}, \citenamefont {Nishioka},\ and\ \citenamefont {Weisbuch}}]{norris1995m}%
  \BibitemOpen
  \bibfield  {author} {\bibinfo {author} {\bibfnamefont {T.~B.}\ \bibnamefont {Norris}}, \bibinfo {author} {\bibfnamefont {J.-K.}\ \bibnamefont {Rhee}}, \bibinfo {author} {\bibfnamefont {C.-Y.}\ \bibnamefont {Sung}}, \bibinfo {author} {\bibfnamefont {Y.}~\bibnamefont {Arakawa}}, \bibinfo {author} {\bibfnamefont {M.}~\bibnamefont {Nishioka}},\ and\ \bibinfo {author} {\bibfnamefont {C.}~\bibnamefont {Weisbuch}},\ }\href {https://doi.org/10.1103/PhysRevB.50.14663} {\bibfield  {journal} {\bibinfo  {journal} {Phys. Rev. B}\ }\textbf {\bibinfo {volume} {50}},\ \bibinfo {pages} {14663} (\bibinfo {year} {1994})}\BibitemShut {NoStop}%
\bibitem [{\citenamefont {Cao}\ \emph {et~al.}(1995)\citenamefont {Cao}, \citenamefont {Jacobson}, \citenamefont {Björk}, \citenamefont {Pau},\ and\ \citenamefont {Yamamoto}}]{10.1063/1.113827}%
  \BibitemOpen
  \bibfield  {author} {\bibinfo {author} {\bibfnamefont {H.}~\bibnamefont {Cao}}, \bibinfo {author} {\bibfnamefont {J.}~\bibnamefont {Jacobson}}, \bibinfo {author} {\bibfnamefont {G.}~\bibnamefont {Björk}}, \bibinfo {author} {\bibfnamefont {S.}~\bibnamefont {Pau}},\ and\ \bibinfo {author} {\bibfnamefont {Y.}~\bibnamefont {Yamamoto}},\ }\href {https://doi.org/10.1063/1.113827} {\bibfield  {journal} {\bibinfo  {journal} {Applied Physics Letters}\ }\textbf {\bibinfo {volume} {66}},\ \bibinfo {pages} {1107} (\bibinfo {year} {1995})},\ \Eprint {https://arxiv.org/abs/https://pubs.aip.org/aip/apl/article-pdf/66/9/1107/18509783/1107\_1\_online.pdf} {https://pubs.aip.org/aip/apl/article-pdf/66/9/1107/18509783/1107\_1\_online.pdf} \BibitemShut {NoStop}%
\bibitem [{\citenamefont {Hennessy}\ \emph {et~al.}(2007)\citenamefont {Hennessy}, \citenamefont {Badolato}, \citenamefont {Winger}, \citenamefont {Gerace}, \citenamefont {Atat\"{u}re}, \citenamefont {Gulde}, \citenamefont {F\"{a}lt}, \citenamefont {Hu},\ and\ \citenamefont {Imamoğlu}}]{Hennessy2007}%
  \BibitemOpen
  \bibfield  {author} {\bibinfo {author} {\bibfnamefont {K.}~\bibnamefont {Hennessy}}, \bibinfo {author} {\bibfnamefont {A.}~\bibnamefont {Badolato}}, \bibinfo {author} {\bibfnamefont {M.}~\bibnamefont {Winger}}, \bibinfo {author} {\bibfnamefont {D.}~\bibnamefont {Gerace}}, \bibinfo {author} {\bibfnamefont {M.}~\bibnamefont {Atat\"{u}re}}, \bibinfo {author} {\bibfnamefont {S.}~\bibnamefont {Gulde}}, \bibinfo {author} {\bibfnamefont {S.}~\bibnamefont {F\"{a}lt}}, \bibinfo {author} {\bibfnamefont {E.~L.}\ \bibnamefont {Hu}},\ and\ \bibinfo {author} {\bibfnamefont {A.}~\bibnamefont {Imamoğlu}},\ }\href {https://doi.org/10.1038/nature05586} {\bibfield  {journal} {\bibinfo  {journal} {Nature}\ }\textbf {\bibinfo {volume} {445}},\ \bibinfo {pages} {896–899} (\bibinfo {year} {2007})}\BibitemShut {NoStop}%
\bibitem [{\citenamefont {Dominici}\ \emph {et~al.}(2021)\citenamefont {Dominici}, \citenamefont {Colas}, \citenamefont {Gianfrate}, \citenamefont {Rahmani}, \citenamefont {Ardizzone}, \citenamefont {Ballarini}, \citenamefont {De~Giorgi}, \citenamefont {Gigli}, \citenamefont {Laussy}, \citenamefont {Sanvitto},\ and\ \citenamefont {Voronova}}]{PhysRevResearch.3.013007}%
  \BibitemOpen
  \bibfield  {author} {\bibinfo {author} {\bibfnamefont {L.}~\bibnamefont {Dominici}}, \bibinfo {author} {\bibfnamefont {D.}~\bibnamefont {Colas}}, \bibinfo {author} {\bibfnamefont {A.}~\bibnamefont {Gianfrate}}, \bibinfo {author} {\bibfnamefont {A.}~\bibnamefont {Rahmani}}, \bibinfo {author} {\bibfnamefont {V.}~\bibnamefont {Ardizzone}}, \bibinfo {author} {\bibfnamefont {D.}~\bibnamefont {Ballarini}}, \bibinfo {author} {\bibfnamefont {M.}~\bibnamefont {De~Giorgi}}, \bibinfo {author} {\bibfnamefont {G.}~\bibnamefont {Gigli}}, \bibinfo {author} {\bibfnamefont {F.~P.}\ \bibnamefont {Laussy}}, \bibinfo {author} {\bibfnamefont {D.}~\bibnamefont {Sanvitto}},\ and\ \bibinfo {author} {\bibfnamefont {N.}~\bibnamefont {Voronova}},\ }\href {https://doi.org/10.1103/PhysRevResearch.3.013007} {\bibfield  {journal} {\bibinfo  {journal} {Phys. Rev. Res.}\ }\textbf {\bibinfo {volume} {3}},\ \bibinfo {pages} {013007} (\bibinfo {year} {2021})}\BibitemShut {NoStop}%
\bibitem [{\citenamefont {Renucci}\ \emph {et~al.}(2005)\citenamefont {Renucci}, \citenamefont {Amand}, \citenamefont {Marie}, \citenamefont {Senellart}, \citenamefont {Bloch}, \citenamefont {Sermage},\ and\ \citenamefont {Kavokin}}]{PhysRevB.72.075317}%
  \BibitemOpen
  \bibfield  {author} {\bibinfo {author} {\bibfnamefont {P.}~\bibnamefont {Renucci}}, \bibinfo {author} {\bibfnamefont {T.}~\bibnamefont {Amand}}, \bibinfo {author} {\bibfnamefont {X.}~\bibnamefont {Marie}}, \bibinfo {author} {\bibfnamefont {P.}~\bibnamefont {Senellart}}, \bibinfo {author} {\bibfnamefont {J.}~\bibnamefont {Bloch}}, \bibinfo {author} {\bibfnamefont {B.}~\bibnamefont {Sermage}},\ and\ \bibinfo {author} {\bibfnamefont {K.~V.}\ \bibnamefont {Kavokin}},\ }\href {https://doi.org/10.1103/PhysRevB.72.075317} {\bibfield  {journal} {\bibinfo  {journal} {Phys. Rev. B}\ }\textbf {\bibinfo {volume} {72}},\ \bibinfo {pages} {075317} (\bibinfo {year} {2005})}\BibitemShut {NoStop}%
\bibitem [{\citenamefont {Mukherjee}\ \emph {et~al.}(2019)\citenamefont {Mukherjee}, \citenamefont {Myers}, \citenamefont {Lena}, \citenamefont {Ozden}, \citenamefont {Beaumariage}, \citenamefont {Sun}, \citenamefont {Steger}, \citenamefont {Pfeiffer}, \citenamefont {West}, \citenamefont {Daley},\ and\ \citenamefont {Snoke}}]{PhysRevB.100.245304}%
  \BibitemOpen
  \bibfield  {author} {\bibinfo {author} {\bibfnamefont {S.}~\bibnamefont {Mukherjee}}, \bibinfo {author} {\bibfnamefont {D.~M.}\ \bibnamefont {Myers}}, \bibinfo {author} {\bibfnamefont {R.~G.}\ \bibnamefont {Lena}}, \bibinfo {author} {\bibfnamefont {B.}~\bibnamefont {Ozden}}, \bibinfo {author} {\bibfnamefont {J.}~\bibnamefont {Beaumariage}}, \bibinfo {author} {\bibfnamefont {Z.}~\bibnamefont {Sun}}, \bibinfo {author} {\bibfnamefont {M.}~\bibnamefont {Steger}}, \bibinfo {author} {\bibfnamefont {L.~N.}\ \bibnamefont {Pfeiffer}}, \bibinfo {author} {\bibfnamefont {K.}~\bibnamefont {West}}, \bibinfo {author} {\bibfnamefont {A.~J.}\ \bibnamefont {Daley}},\ and\ \bibinfo {author} {\bibfnamefont {D.~W.}\ \bibnamefont {Snoke}},\ }\href {https://doi.org/10.1103/PhysRevB.100.245304} {\bibfield  {journal} {\bibinfo  {journal} {Phys. Rev. B}\ }\textbf {\bibinfo {volume} {100}},\ \bibinfo {pages} {245304} (\bibinfo {year} {2019})}\BibitemShut {NoStop}%
\bibitem [{\citenamefont {Kim}\ \emph {et~al.}(2020)\citenamefont {Kim}, \citenamefont {Rubo}, \citenamefont {Liew}, \citenamefont {Brodbeck}, \citenamefont {Schneider}, \citenamefont {H\"ofling},\ and\ \citenamefont {Deng}}]{PhysRevB.101.085302}%
  \BibitemOpen
  \bibfield  {author} {\bibinfo {author} {\bibfnamefont {S.}~\bibnamefont {Kim}}, \bibinfo {author} {\bibfnamefont {Y.~G.}\ \bibnamefont {Rubo}}, \bibinfo {author} {\bibfnamefont {T.~C.~H.}\ \bibnamefont {Liew}}, \bibinfo {author} {\bibfnamefont {S.}~\bibnamefont {Brodbeck}}, \bibinfo {author} {\bibfnamefont {C.}~\bibnamefont {Schneider}}, \bibinfo {author} {\bibfnamefont {S.}~\bibnamefont {H\"ofling}},\ and\ \bibinfo {author} {\bibfnamefont {H.}~\bibnamefont {Deng}},\ }\href {https://doi.org/10.1103/PhysRevB.101.085302} {\bibfield  {journal} {\bibinfo  {journal} {Phys. Rev. B}\ }\textbf {\bibinfo {volume} {101}},\ \bibinfo {pages} {085302} (\bibinfo {year} {2020})}\BibitemShut {NoStop}%
\bibitem [{\citenamefont {Rayanov}\ \emph {et~al.}(2015)\citenamefont {Rayanov}, \citenamefont {Altshuler}, \citenamefont {Rubo},\ and\ \citenamefont {Flach}}]{PhysRevLett.114.193901}%
  \BibitemOpen
  \bibfield  {author} {\bibinfo {author} {\bibfnamefont {K.}~\bibnamefont {Rayanov}}, \bibinfo {author} {\bibfnamefont {B.~L.}\ \bibnamefont {Altshuler}}, \bibinfo {author} {\bibfnamefont {Y.~G.}\ \bibnamefont {Rubo}},\ and\ \bibinfo {author} {\bibfnamefont {S.}~\bibnamefont {Flach}},\ }\href {https://doi.org/10.1103/PhysRevLett.114.193901} {\bibfield  {journal} {\bibinfo  {journal} {Phys. Rev. Lett.}\ }\textbf {\bibinfo {volume} {114}},\ \bibinfo {pages} {193901} (\bibinfo {year} {2015})}\BibitemShut {NoStop}%
\bibitem [{\citenamefont {Nalitov}\ \emph {et~al.}(2019)\citenamefont {Nalitov}, \citenamefont {Sigurdsson}, \citenamefont {Morina}, \citenamefont {Krivosenko}, \citenamefont {Iorsh}, \citenamefont {Rubo}, \citenamefont {Kavokin},\ and\ \citenamefont {Shelykh}}]{PhysRevA.99.033830}%
  \BibitemOpen
  \bibfield  {author} {\bibinfo {author} {\bibfnamefont {A.~V.}\ \bibnamefont {Nalitov}}, \bibinfo {author} {\bibfnamefont {H.}~\bibnamefont {Sigurdsson}}, \bibinfo {author} {\bibfnamefont {S.}~\bibnamefont {Morina}}, \bibinfo {author} {\bibfnamefont {Y.~S.}\ \bibnamefont {Krivosenko}}, \bibinfo {author} {\bibfnamefont {I.~V.}\ \bibnamefont {Iorsh}}, \bibinfo {author} {\bibfnamefont {Y.~G.}\ \bibnamefont {Rubo}}, \bibinfo {author} {\bibfnamefont {A.~V.}\ \bibnamefont {Kavokin}},\ and\ \bibinfo {author} {\bibfnamefont {I.~A.}\ \bibnamefont {Shelykh}},\ }\href {https://doi.org/10.1103/PhysRevA.99.033830} {\bibfield  {journal} {\bibinfo  {journal} {Phys. Rev. A}\ }\textbf {\bibinfo {volume} {99}},\ \bibinfo {pages} {033830} (\bibinfo {year} {2019})}\BibitemShut {NoStop}%
\bibitem [{\citenamefont {Sigurdsson}\ \emph {et~al.}(2022)\citenamefont {Sigurdsson}, \citenamefont {Gnusov}, \citenamefont {Alyatkin}, \citenamefont {Pickup}, \citenamefont {Gippius}, \citenamefont {Lagoudakis},\ and\ \citenamefont {Askitopoulos}}]{PhysRevLett.129.155301}%
  \BibitemOpen
  \bibfield  {author} {\bibinfo {author} {\bibfnamefont {H.}~\bibnamefont {Sigurdsson}}, \bibinfo {author} {\bibfnamefont {I.}~\bibnamefont {Gnusov}}, \bibinfo {author} {\bibfnamefont {S.}~\bibnamefont {Alyatkin}}, \bibinfo {author} {\bibfnamefont {L.}~\bibnamefont {Pickup}}, \bibinfo {author} {\bibfnamefont {N.~A.}\ \bibnamefont {Gippius}}, \bibinfo {author} {\bibfnamefont {P.~G.}\ \bibnamefont {Lagoudakis}},\ and\ \bibinfo {author} {\bibfnamefont {A.}~\bibnamefont {Askitopoulos}},\ }\href {https://doi.org/10.1103/PhysRevLett.129.155301} {\bibfield  {journal} {\bibinfo  {journal} {Phys. Rev. Lett.}\ }\textbf {\bibinfo {volume} {129}},\ \bibinfo {pages} {155301} (\bibinfo {year} {2022})}\BibitemShut {NoStop}%
\bibitem [{\citenamefont {Barrat}\ \emph {et~al.}(2023)\citenamefont {Barrat}, \citenamefont {Tzortzakakis}, \citenamefont {Niu}, \citenamefont {Zhou}, \citenamefont {Paschos}, \citenamefont {Petrosyan},\ and\ \citenamefont {Savvidis}}]{barrat2023superfluid}%
  \BibitemOpen
  \bibfield  {author} {\bibinfo {author} {\bibfnamefont {J.}~\bibnamefont {Barrat}}, \bibinfo {author} {\bibfnamefont {A.}~\bibnamefont {Tzortzakakis}}, \bibinfo {author} {\bibfnamefont {M.}~\bibnamefont {Niu}}, \bibinfo {author} {\bibfnamefont {X.}~\bibnamefont {Zhou}}, \bibinfo {author} {\bibfnamefont {G.}~\bibnamefont {Paschos}}, \bibinfo {author} {\bibfnamefont {D.}~\bibnamefont {Petrosyan}},\ and\ \bibinfo {author} {\bibfnamefont {P.}~\bibnamefont {Savvidis}},\ }\bibfield  {journal} {\bibinfo  {journal} {arXiv preprint arXiv:2308.05555}\ }\href {https://doi.org/https://doi.org/10.48550/arXiv.2308.05555} {https://doi.org/10.48550/arXiv.2308.05555} (\bibinfo {year} {2023})\BibitemShut {NoStop}%
\bibitem [{\citenamefont {Carraro-Haddad}\ \emph {et~al.}(2024)\citenamefont {Carraro-Haddad}, \citenamefont {Chafatinos}, \citenamefont {Kuznetsov}, \citenamefont {Papuccio-Fern\'andez}, \citenamefont {Reynoso}, \citenamefont {Bruchhausen}, \citenamefont {Biermann}, \citenamefont {Santos}, \citenamefont {Usaj},\ and\ \citenamefont {Fainstein}}]{CarraroHaddad2024}%
  \BibitemOpen
  \bibfield  {author} {\bibinfo {author} {\bibfnamefont {I.}~\bibnamefont {Carraro-Haddad}}, \bibinfo {author} {\bibfnamefont {D.~L.}\ \bibnamefont {Chafatinos}}, \bibinfo {author} {\bibfnamefont {A.~S.}\ \bibnamefont {Kuznetsov}}, \bibinfo {author} {\bibfnamefont {I.~A.}\ \bibnamefont {Papuccio-Fern\'andez}}, \bibinfo {author} {\bibfnamefont {A.~A.}\ \bibnamefont {Reynoso}}, \bibinfo {author} {\bibfnamefont {A.}~\bibnamefont {Bruchhausen}}, \bibinfo {author} {\bibfnamefont {K.}~\bibnamefont {Biermann}}, \bibinfo {author} {\bibfnamefont {P.~V.}\ \bibnamefont {Santos}}, \bibinfo {author} {\bibfnamefont {G.}~\bibnamefont {Usaj}},\ and\ \bibinfo {author} {\bibfnamefont {A.}~\bibnamefont {Fainstein}},\ }\href {https://doi.org/10.1126/science.adn7087} {\bibfield  {journal} {\bibinfo  {journal} {Science}\ }\textbf {\bibinfo {volume} {384}},\ \bibinfo {pages} {995} (\bibinfo {year} {2024})}\BibitemShut {NoStop}%
\bibitem [{\citenamefont {Sun}\ \emph {et~al.}(2024)\citenamefont {Sun}, \citenamefont {Wang}, \citenamefont {Hou}, \citenamefont {Bi}, \citenamefont {Xue},\ and\ \citenamefont {Kavokin}}]{PhysRevB.109.155301}%
  \BibitemOpen
  \bibfield  {author} {\bibinfo {author} {\bibfnamefont {X.}~\bibnamefont {Sun}}, \bibinfo {author} {\bibfnamefont {G.}~\bibnamefont {Wang}}, \bibinfo {author} {\bibfnamefont {K.}~\bibnamefont {Hou}}, \bibinfo {author} {\bibfnamefont {H.}~\bibnamefont {Bi}}, \bibinfo {author} {\bibfnamefont {Y.}~\bibnamefont {Xue}},\ and\ \bibinfo {author} {\bibfnamefont {A.}~\bibnamefont {Kavokin}},\ }\href {https://doi.org/10.1103/PhysRevB.109.155301} {\bibfield  {journal} {\bibinfo  {journal} {Phys. Rev. B}\ }\textbf {\bibinfo {volume} {109}},\ \bibinfo {pages} {155301} (\bibinfo {year} {2024})}\BibitemShut {NoStop}%
\bibitem [{\citenamefont {Aladinskaia}\ \emph {et~al.}(2023)\citenamefont {Aladinskaia}, \citenamefont {Cherbunin}, \citenamefont {Sedov}, \citenamefont {Liubomirov}, \citenamefont {Kavokin}, \citenamefont {Khramtsov}, \citenamefont {Petrov}, \citenamefont {Savvidis},\ and\ \citenamefont {Kavokin}}]{Aladinskaia}%
  \BibitemOpen
  \bibfield  {author} {\bibinfo {author} {\bibfnamefont {E.}~\bibnamefont {Aladinskaia}}, \bibinfo {author} {\bibfnamefont {R.}~\bibnamefont {Cherbunin}}, \bibinfo {author} {\bibfnamefont {E.}~\bibnamefont {Sedov}}, \bibinfo {author} {\bibfnamefont {A.}~\bibnamefont {Liubomirov}}, \bibinfo {author} {\bibfnamefont {K.}~\bibnamefont {Kavokin}}, \bibinfo {author} {\bibfnamefont {E.}~\bibnamefont {Khramtsov}}, \bibinfo {author} {\bibfnamefont {M.}~\bibnamefont {Petrov}}, \bibinfo {author} {\bibfnamefont {P.}~\bibnamefont {Savvidis}},\ and\ \bibinfo {author} {\bibfnamefont {A.}~\bibnamefont {Kavokin}},\ }\href {https://doi.org/10.1103/PhysRevB.107.045302} {\bibfield  {journal} {\bibinfo  {journal} {Phys. Rev. B}\ }\textbf {\bibinfo {volume} {107}},\ \bibinfo {pages} {045302} (\bibinfo {year} {2023})}\BibitemShut {NoStop}%
\bibitem [{\citenamefont {Yao}\ \emph {et~al.}(2023)\citenamefont {Yao}, \citenamefont {Comaron}, \citenamefont {Alnatah}, \citenamefont {Beaumariage}, \citenamefont {Mukherjee}, \citenamefont {West}, \citenamefont {Pfeiffer}, \citenamefont {Baldwin}, \citenamefont {Szymanska},\ and\ \citenamefont {Snoke}}]{Snoke}%
  \BibitemOpen
  \bibfield  {author} {\bibinfo {author} {\bibfnamefont {Q.}~\bibnamefont {Yao}}, \bibinfo {author} {\bibfnamefont {P.}~\bibnamefont {Comaron}}, \bibinfo {author} {\bibfnamefont {H.}~\bibnamefont {Alnatah}}, \bibinfo {author} {\bibfnamefont {J.}~\bibnamefont {Beaumariage}}, \bibinfo {author} {\bibfnamefont {S.}~\bibnamefont {Mukherjee}}, \bibinfo {author} {\bibfnamefont {K.}~\bibnamefont {West}}, \bibinfo {author} {\bibfnamefont {L.}~\bibnamefont {Pfeiffer}}, \bibinfo {author} {\bibfnamefont {K.}~\bibnamefont {Baldwin}}, \bibinfo {author} {\bibfnamefont {M.}~\bibnamefont {Szymanska}},\ and\ \bibinfo {author} {\bibfnamefont {D.}~\bibnamefont {Snoke}},\ }\href {https://doi.org/10.48550/arXiv.2302.07803} {\bibinfo {title} {Persistent, controllable circulation of a polariton ring condensate}},\ \bibinfo {howpublished} {arXiv preprint arXiv:2302.07803} (\bibinfo {year} {2023})\BibitemShut {NoStop}%
\bibitem [{\citenamefont {T\"opfer}\ \emph {et~al.}(2020)\citenamefont {T\"opfer}, \citenamefont {Sigurdsson}, \citenamefont {Alyatkin},\ and\ \citenamefont {Lagoudakis}}]{PhysRevB.102.195428}%
  \BibitemOpen
  \bibfield  {author} {\bibinfo {author} {\bibfnamefont {J.~D.}\ \bibnamefont {T\"opfer}}, \bibinfo {author} {\bibfnamefont {H.}~\bibnamefont {Sigurdsson}}, \bibinfo {author} {\bibfnamefont {S.}~\bibnamefont {Alyatkin}},\ and\ \bibinfo {author} {\bibfnamefont {P.~G.}\ \bibnamefont {Lagoudakis}},\ }\href {https://doi.org/10.1103/PhysRevB.102.195428} {\bibfield  {journal} {\bibinfo  {journal} {Phys. Rev. B}\ }\textbf {\bibinfo {volume} {102}},\ \bibinfo {pages} {195428} (\bibinfo {year} {2020})}\BibitemShut {NoStop}%
\bibitem [{\citenamefont {Estrecho}\ \emph {et~al.}(2021)\citenamefont {Estrecho}, \citenamefont {Pieczarka}, \citenamefont {Wurdack}, \citenamefont {Steger}, \citenamefont {West}, \citenamefont {Pfeiffer}, \citenamefont {Snoke}, \citenamefont {Truscott},\ and\ \citenamefont {Ostrovskaya}}]{PhysRevLett.126.075301}%
  \BibitemOpen
  \bibfield  {author} {\bibinfo {author} {\bibfnamefont {E.}~\bibnamefont {Estrecho}}, \bibinfo {author} {\bibfnamefont {M.}~\bibnamefont {Pieczarka}}, \bibinfo {author} {\bibfnamefont {M.}~\bibnamefont {Wurdack}}, \bibinfo {author} {\bibfnamefont {M.}~\bibnamefont {Steger}}, \bibinfo {author} {\bibfnamefont {K.}~\bibnamefont {West}}, \bibinfo {author} {\bibfnamefont {L.~N.}\ \bibnamefont {Pfeiffer}}, \bibinfo {author} {\bibfnamefont {D.~W.}\ \bibnamefont {Snoke}}, \bibinfo {author} {\bibfnamefont {A.~G.}\ \bibnamefont {Truscott}},\ and\ \bibinfo {author} {\bibfnamefont {E.~A.}\ \bibnamefont {Ostrovskaya}},\ }\href {https://doi.org/10.1103/PhysRevLett.126.075301} {\bibfield  {journal} {\bibinfo  {journal} {Phys. Rev. Lett.}\ }\textbf {\bibinfo {volume} {126}},\ \bibinfo {pages} {075301} (\bibinfo {year} {2021})}\BibitemShut {NoStop}%
\bibitem [{\citenamefont {Xue}\ \emph {et~al.}(2021)\citenamefont {Xue}, \citenamefont {Chestnov}, \citenamefont {Sedov}, \citenamefont {Kiktenko}, \citenamefont {Fedorov}, \citenamefont {Schumacher}, \citenamefont {Ma},\ and\ \citenamefont {Kavokin}}]{PhysRevResearch.3.013099}%
  \BibitemOpen
  \bibfield  {author} {\bibinfo {author} {\bibfnamefont {Y.}~\bibnamefont {Xue}}, \bibinfo {author} {\bibfnamefont {I.}~\bibnamefont {Chestnov}}, \bibinfo {author} {\bibfnamefont {E.}~\bibnamefont {Sedov}}, \bibinfo {author} {\bibfnamefont {E.}~\bibnamefont {Kiktenko}}, \bibinfo {author} {\bibfnamefont {A.~K.}\ \bibnamefont {Fedorov}}, \bibinfo {author} {\bibfnamefont {S.}~\bibnamefont {Schumacher}}, \bibinfo {author} {\bibfnamefont {X.}~\bibnamefont {Ma}},\ and\ \bibinfo {author} {\bibfnamefont {A.}~\bibnamefont {Kavokin}},\ }\href {https://doi.org/10.1103/PhysRevResearch.3.013099} {\bibfield  {journal} {\bibinfo  {journal} {Phys. Rev. Res.}\ }\textbf {\bibinfo {volume} {3}},\ \bibinfo {pages} {013099} (\bibinfo {year} {2021})}\BibitemShut {NoStop}%
\bibitem [{\citenamefont {Kavokin}\ \emph {et~al.}(2022)\citenamefont {Kavokin}, \citenamefont {Liew}, \citenamefont {Schneider}, \citenamefont {Lagoudakis}, \citenamefont {Klembt},\ and\ \citenamefont {Hoefling}}]{Kavokin2022}%
  \BibitemOpen
  \bibfield  {author} {\bibinfo {author} {\bibfnamefont {A.}~\bibnamefont {Kavokin}}, \bibinfo {author} {\bibfnamefont {T.~C.~H.}\ \bibnamefont {Liew}}, \bibinfo {author} {\bibfnamefont {C.}~\bibnamefont {Schneider}}, \bibinfo {author} {\bibfnamefont {P.~G.}\ \bibnamefont {Lagoudakis}}, \bibinfo {author} {\bibfnamefont {S.}~\bibnamefont {Klembt}},\ and\ \bibinfo {author} {\bibfnamefont {S.}~\bibnamefont {Hoefling}},\ }\href {https://doi.org/10.1038/s42254-022-00447-1} {\bibfield  {journal} {\bibinfo  {journal} {Nature Reviews Physics}\ }\textbf {\bibinfo {volume} {4}},\ \bibinfo {pages} {435–451} (\bibinfo {year} {2022})}\BibitemShut {NoStop}%
\bibitem [{\citenamefont {Ricco}\ \emph {et~al.}(2024)\citenamefont {Ricco}, \citenamefont {Shelykh},\ and\ \citenamefont {Kavokin}}]{Ricco2024}%
  \BibitemOpen
  \bibfield  {author} {\bibinfo {author} {\bibfnamefont {L.~S.}\ \bibnamefont {Ricco}}, \bibinfo {author} {\bibfnamefont {I.~A.}\ \bibnamefont {Shelykh}},\ and\ \bibinfo {author} {\bibfnamefont {A.}~\bibnamefont {Kavokin}},\ }\bibfield  {journal} {\bibinfo  {journal} {Scientific Reports}\ }\textbf {\bibinfo {volume} {14}},\ \href {https://doi.org/10.1038/s41598-024-54543-6} {10.1038/s41598-024-54543-6} (\bibinfo {year} {2024})\BibitemShut {NoStop}%
\bibitem [{\citenamefont {Barrat}\ \emph {et~al.}(2024)\citenamefont {Barrat} \emph {et~al.}}]{Barrat2024}%
  \BibitemOpen
  \bibfield  {author} {\bibinfo {author} {\bibfnamefont {J.}~\bibnamefont {Barrat}} \emph {et~al.},\ }\href@noop {} {\bibfield  {journal} {\bibinfo  {journal} {Science Advances}\ }\textbf {\bibinfo {volume} {10}},\ \bibinfo {pages} {eado4042} (\bibinfo {year} {2024})}\BibitemShut {NoStop}%
\bibitem [{\citenamefont {de~Oliveira}\ and\ \citenamefont {Munro}(2000)}]{deOliveira2000}%
  \BibitemOpen
  \bibfield  {author} {\bibinfo {author} {\bibfnamefont {M.}~\bibnamefont {de~Oliveira}}\ and\ \bibinfo {author} {\bibfnamefont {W.~J.}\ \bibnamefont {Munro}},\ }\href@noop {} {\bibfield  {journal} {\bibinfo  {journal} {Physical Review A}\ }\textbf {\bibinfo {volume} {61}},\ \bibinfo {pages} {042309} (\bibinfo {year} {2000})}\BibitemShut {NoStop}%
\bibitem [{\citenamefont {Kiktenko}\ \emph {et~al.}(2015)\citenamefont {Kiktenko}, \citenamefont {Fedorov}, \citenamefont {Man'ko},\ and\ \citenamefont {Man'ko}}]{Kiktenko2015}%
  \BibitemOpen
  \bibfield  {author} {\bibinfo {author} {\bibfnamefont {E.}~\bibnamefont {Kiktenko}}, \bibinfo {author} {\bibfnamefont {A.}~\bibnamefont {Fedorov}}, \bibinfo {author} {\bibfnamefont {O.}~\bibnamefont {Man'ko}},\ and\ \bibinfo {author} {\bibfnamefont {V.}~\bibnamefont {Man'ko}},\ }\href@noop {} {\bibfield  {journal} {\bibinfo  {journal} {Physical Review A}\ }\textbf {\bibinfo {volume} {91}},\ \bibinfo {pages} {042312} (\bibinfo {year} {2015})}\BibitemShut {NoStop}%
\bibitem [{\citenamefont {Vidal}\ and\ \citenamefont {Werner}(2002)}]{Vidal2002}%
  \BibitemOpen
  \bibfield  {author} {\bibinfo {author} {\bibfnamefont {G.}~\bibnamefont {Vidal}}\ and\ \bibinfo {author} {\bibfnamefont {R.~F.}\ \bibnamefont {Werner}},\ }\href@noop {} {\bibfield  {journal} {\bibinfo  {journal} {Physical Review A}\ }\textbf {\bibinfo {volume} {65}},\ \bibinfo {pages} {032314} (\bibinfo {year} {2002})}\BibitemShut {NoStop}%
\bibitem [{\citenamefont {Dominici}\ \emph {et~al.}(2014)\citenamefont {Dominici}, \citenamefont {Colas}, \citenamefont {Donati}, \citenamefont {Restrepo~Cuartas}, \citenamefont {De~Giorgi}, \citenamefont {Ballarini}, \citenamefont {Guirales}, \citenamefont {L\'opez Carre\~no}, \citenamefont {Bramati}, \citenamefont {Gigli}, \citenamefont {del Valle}, \citenamefont {Laussy},\ and\ \citenamefont {Sanvitto}}]{PhysRevLett.113.226401}%
  \BibitemOpen
  \bibfield  {author} {\bibinfo {author} {\bibfnamefont {L.}~\bibnamefont {Dominici}}, \bibinfo {author} {\bibfnamefont {D.}~\bibnamefont {Colas}}, \bibinfo {author} {\bibfnamefont {S.}~\bibnamefont {Donati}}, \bibinfo {author} {\bibfnamefont {J.~P.}\ \bibnamefont {Restrepo~Cuartas}}, \bibinfo {author} {\bibfnamefont {M.}~\bibnamefont {De~Giorgi}}, \bibinfo {author} {\bibfnamefont {D.}~\bibnamefont {Ballarini}}, \bibinfo {author} {\bibfnamefont {G.}~\bibnamefont {Guirales}}, \bibinfo {author} {\bibfnamefont {J.~C.}\ \bibnamefont {L\'opez Carre\~no}}, \bibinfo {author} {\bibfnamefont {A.}~\bibnamefont {Bramati}}, \bibinfo {author} {\bibfnamefont {G.}~\bibnamefont {Gigli}}, \bibinfo {author} {\bibfnamefont {E.}~\bibnamefont {del Valle}}, \bibinfo {author} {\bibfnamefont {F.~P.}\ \bibnamefont {Laussy}},\ and\ \bibinfo {author} {\bibfnamefont {D.}~\bibnamefont {Sanvitto}},\ }\href {https://doi.org/10.1103/PhysRevLett.113.226401} {\bibfield  {journal} {\bibinfo  {journal} {Phys. Rev. Lett.}\ }\textbf {\bibinfo
  {volume} {113}},\ \bibinfo {pages} {226401} (\bibinfo {year} {2014})}\BibitemShut {NoStop}%
\bibitem [{\citenamefont {Faust}\ \emph {et~al.}(2013)\citenamefont {Faust}, \citenamefont {Rieger}, \citenamefont {Seitner}, \citenamefont {Kotthaus},\ and\ \citenamefont {Weig}}]{Faust2013}%
  \BibitemOpen
  \bibfield  {author} {\bibinfo {author} {\bibfnamefont {T.}~\bibnamefont {Faust}}, \bibinfo {author} {\bibfnamefont {J.}~\bibnamefont {Rieger}}, \bibinfo {author} {\bibfnamefont {M.~J.}\ \bibnamefont {Seitner}}, \bibinfo {author} {\bibfnamefont {J.~P.}\ \bibnamefont {Kotthaus}},\ and\ \bibinfo {author} {\bibfnamefont {E.~M.}\ \bibnamefont {Weig}},\ }\href {https://doi.org/10.1038/nphys2666} {\bibfield  {journal} {\bibinfo  {journal} {Nature Physics}\ }\textbf {\bibinfo {volume} {9}},\ \bibinfo {pages} {485–488} (\bibinfo {year} {2013})}\BibitemShut {NoStop}%
\bibitem [{\citenamefont {Frimmer}\ and\ \citenamefont {Novotny}(2014)}]{10.1119/1.4878621}%
  \BibitemOpen
  \bibfield  {author} {\bibinfo {author} {\bibfnamefont {M.}~\bibnamefont {Frimmer}}\ and\ \bibinfo {author} {\bibfnamefont {L.}~\bibnamefont {Novotny}},\ }\href {https://doi.org/10.1119/1.4878621} {\bibfield  {journal} {\bibinfo  {journal} {American Journal of Physics}\ }\textbf {\bibinfo {volume} {82}},\ \bibinfo {pages} {947} (\bibinfo {year} {2014})},\ \Eprint {https://arxiv.org/abs/https://pubs.aip.org/aapt/ajp/article-pdf/82/10/947/12832121/947\_1\_online.pdf} {https://pubs.aip.org/aapt/ajp/article-pdf/82/10/947/12832121/947\_1\_online.pdf} \BibitemShut {NoStop}%
\bibitem [{\citenamefont {Snoke}(2024)}]{Snoke2024-ah}%
  \BibitemOpen
  \bibfield  {author} {\bibinfo {author} {\bibfnamefont {D.~W.}\ \bibnamefont {Snoke}},\ }\href@noop {} {\emph {\bibinfo {title} {Interpreting quantum mechanics: Modern foundations}}}\ (\bibinfo  {publisher} {Cambridge University Press},\ \bibinfo {address} {Cambridge, England},\ \bibinfo {year} {2024})\ \bibinfo {note} {chapter 21.3}\BibitemShut {NoStop}%
\end{thebibliography}%

\newpage
\clearpage
\begin{widetext}

\renewcommand{\thefigure}{S\arabic{figure}}
\renewcommand{\thetable}{S\arabic{table}}
\renewcommand{\theequation}{S\arabic{equation}}
\renewcommand{\thepage}{S\arabic{page}}
\setcounter{figure}{0}
\setcounter{table}{0}
\setcounter{equation}{0}
\setcounter{page}{1} 


\begin{center}
\section*{Supplementary Materials for Quantum beats of a macroscopic polariton condensate in real space}
R.V. Cherbunin$^{\ast}$,
A. Liubomirov$^{}$,
D. Novokreschenov$^{}$, \and
A. Kudlis$^{}$,
A.V. Kavokin$^{\ast}$, \and 

\small$^\ast$Corresponding author. Email: r.cherbunin@spbu.ru, kavokinalexey@gmail.com\\
\end{center}

\section*{Methods}

The sample under study is a planar structure grown by molecular-beam epitaxy. The bottom Bragg mirror, a cavity with twelve embedded quantum wells and the top Bragg mirror were grown subsequently on a GaAs substrate. The top (bottom) mirror consists of 40(45) pairs of AlAs/GaAlAs $\lambda/4$-layers. The AlGaAs $3\lambda/2$ cavity includes three sets by 4 GaAs quantum wells placed in the anti-nodes of the cavity mode. The quality factor of the sample was 16000. The gradient of cavity width made it possible to select the detuning between quantum well exciton and photon modes by moving the laser spot on the surface of the sample. All experiments were carried out at the small negative detuning. The sample was cooled out to the temperature 6 K using a closed-cycle low-vibration cryostat. The vibration level of the sample was less than 100 nm. 

\begin{figure}[b]
    \centering
    \includegraphics[width=0.90\linewidth]{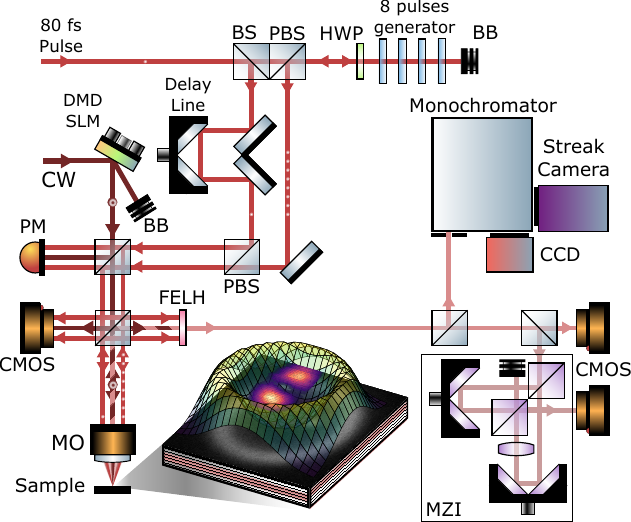}
    \caption{Experimental set-up. BS — non-polarizing beam splitter. PBS — polarizing beam splitter. BB — black body. DMD SLM — spatial light modulator based on digital micromirror device. CW — single-mode {\em cw} laser with $E_{ph} = 1.65$ eV. PM — power meter. MO — micro objective. FELH — low-pass spectral filter. MZI — Mach-Zhender interferometer. }
    \label{Sfig:1}
\end{figure}

\newpage

\begin{figure}[t]
    \centering
    \includegraphics[width=1.0\linewidth]{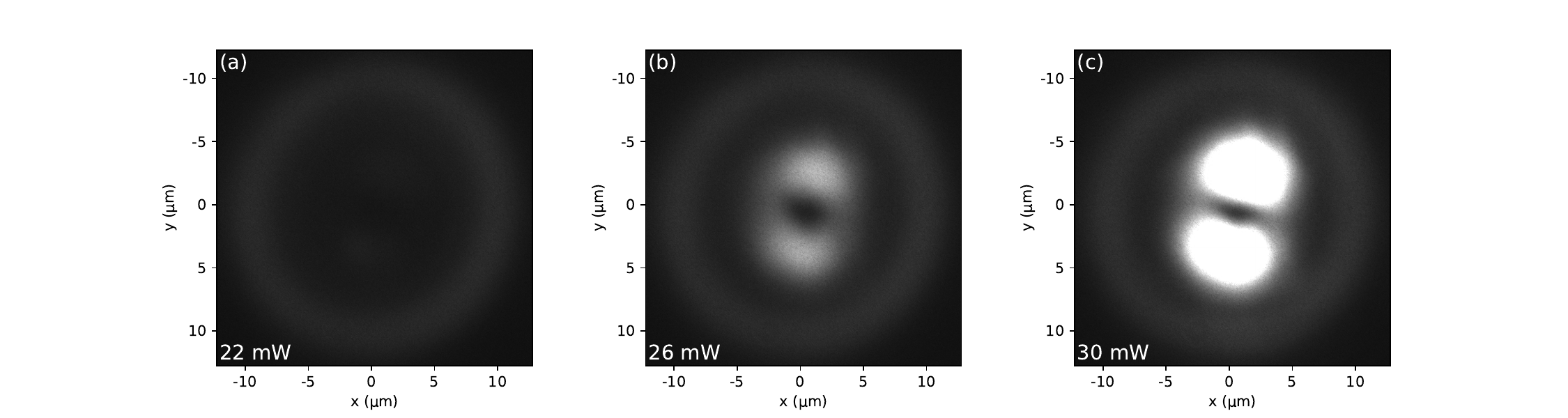}
    \caption{Emission of the polariton laser at different pump powers of the trap. Week external ellipse shows approximate position of the trap while the two bright emitting spots with a hole at the center is a polariton condensate. Trap laser powers are shown as numbers on the images. }
    \label{Sfig:2}
\end{figure}

The laser beam of the {\em{cw}} single-mode semiconductor laser was focused on the sample surface with a 50x objective in the shape of an ellipse. The photon energy of the laser was 1.65 eV corresponding to the first dip in the reflection coefficient of the top Bragg mirror of the sample. The elliptical shape of the laser beam was created using a spatial light modulator (SLM, micro-mirror array with 1M of 16x16 $\mu$m computer-controlled mirrors).  The internal diameter of the resulting trap was about 20 $\mu$m while the width of the trap was about 1 $\mu$m. The trap ellipticity, defined as a ratio of the major to minor axis lengths, was varied in a small range 1 .. 1.1. The width and the diameter of the trap were chosen in such a way that the first excited state of the trap was mostly populated by exciton-polaritons. The corresponding polariton condensate density distribution looks like a dumbbell oriented along the major axis of the trap. The orientation of this dumbbell-like state follows the orientation of the elliptical trap if we rotate it as figure~\ref{Sfig:3} shows. The laser radiation power was controlled using a half-wave phase plate and a linear polarizer within the range of 0-30 mW and was time modulated. The lasing threshold of a polariton laser corresponds to the pump power of the order of 25 mV. The heating of the sample, which was inevitable at such pump power, apparently determined the coherence time of the observed oscillations in our case. 

\begin{figure}[b]
    \centering
    \includegraphics[width=0.90\linewidth]{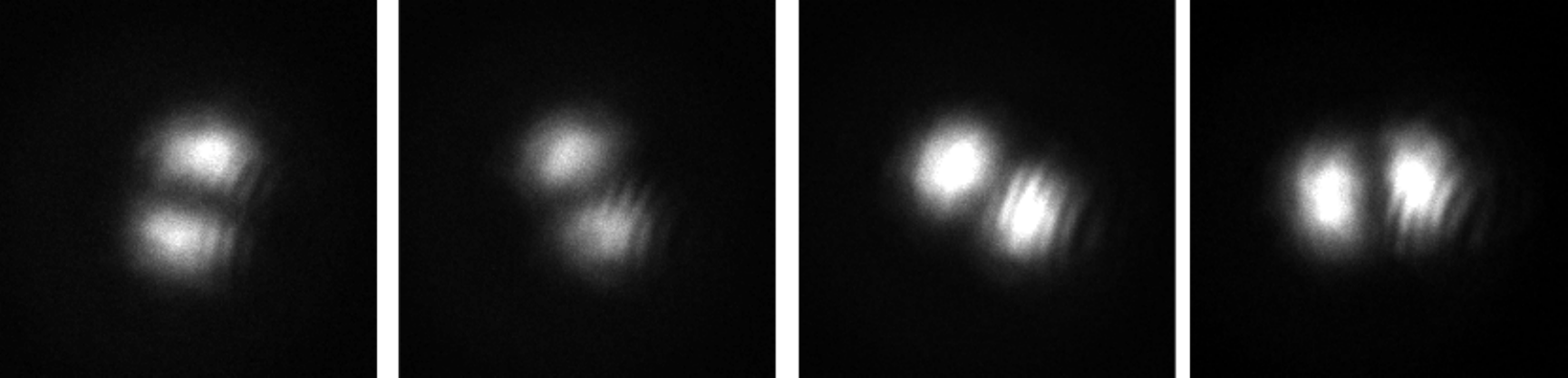}
    \caption{Emission of the polariton condensate at different orientations of the trap. Sequential images correspond to the following orientations of the trap: -10, 20, 45, 90 degrees. }
    \label{Sfig:3}
\end{figure}

\newpage

\begin{figure}[t]
    \centering
    \includegraphics[width=0.90\linewidth]{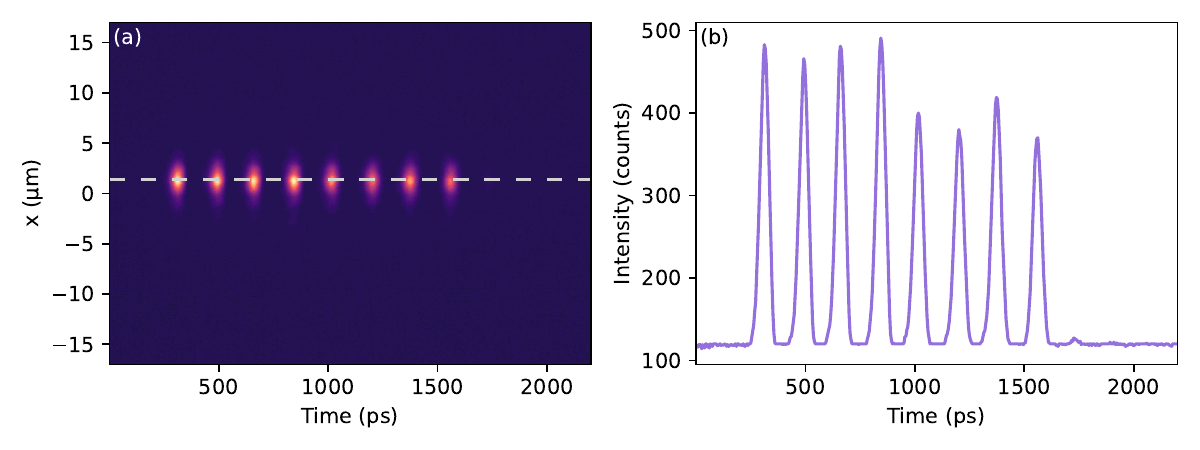}
    \caption{(a) The sequence of 8 laser pulses used for Hadamard operations measured with a streak camera. Dashed white line indicates the cross-section along which the the time dependence at graph (b) is plotted.}
    \label{Sfig:4}
\end{figure}

In addition to {\em{cw}} laser radiation, two beams of a pulsed femtosecond titanium-sapphire laser with the same wavelength were focused at the sample to control the polariton dynamics. The first beam which initiated quantum beats consisted of eight pulses coming one after another at 190 ps intervals. To create such a beam we passed the laser light through four glass plates of equal thicknesses installed parallel to each other at a certain distance. The reflection collected from this set-up represented the pulse train shown in figure~\ref{Sfig:4}. This laser beam was then directed through a 10:90 beam splitter onto the sample surface and served as a control pulse for the Hadamard operation. Another beam passed through a motorized delay line and was also focused onto the sample. Its position on the sample could be changed with respect to the position of the first pulse. The diameter of the control beams on the sample did not exceed 1 micron.

\begin{figure}[b]
    \centering
    \includegraphics[width=1.0\linewidth]{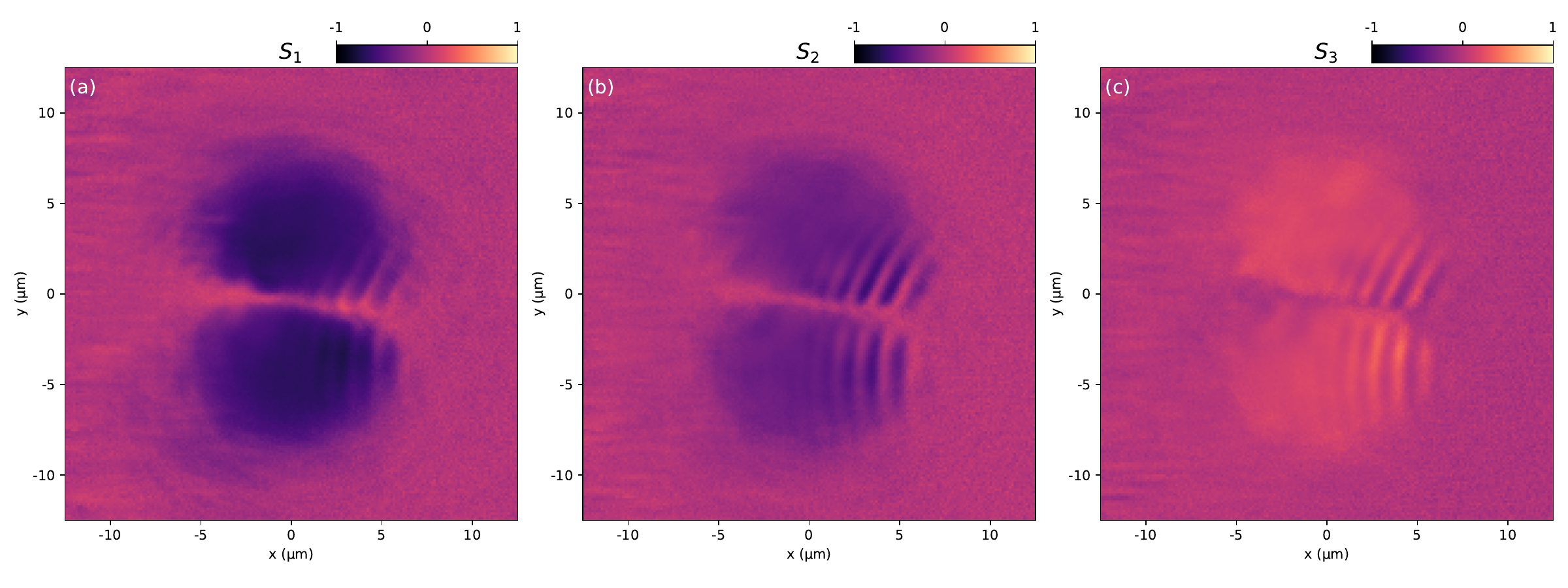}
    \caption{Stokes components of polariton emission. (a) Linear V-H. (b) Linear A-D. (c) Circular.}
    \label{fig:5}
\end{figure}

\newpage

\begin{figure}[t]
    \centering
    \includegraphics[width=1.0\linewidth]{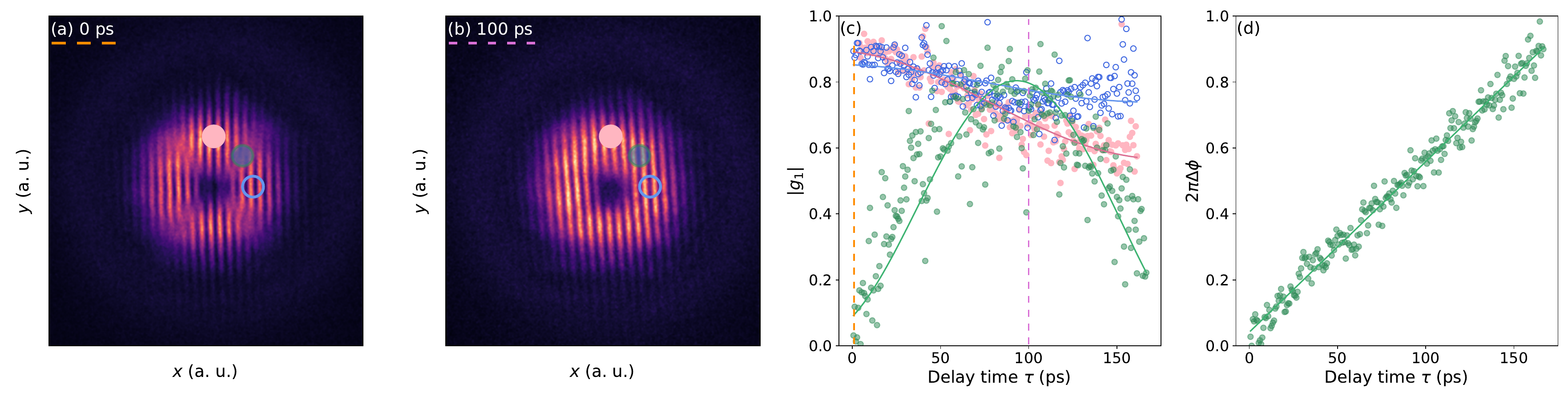}
    \caption{
    The time-resolved measurements of the first-order coherence are presented. Figures (a) and (b) show the interferograms of the condensate recorded at the time delays of 0 and 100\,ps, respectively. The pale pink filled circle marks the $p_y$ orbital (as seen in panels (a) and (b)) with a time delay of $\tau_{p_y} = 753.1$\,ps, while the blue unfilled circle represents the $p_x$ orbital at $\tau_{p_x} = 1217.1$\,ps. The green half-filled circle indicates the region where the orbitals spatially overlap. In panel (c), the resulting first-order correlation function, $\vert g_1(\tau)\vert$, is displayed. The green curve represents a fit of the oscillatory behavior of the $p_x$–$p_y$ superposition coherence using a theoretical two-mode model, yielding an oscillation period $T_{g_1} = 193.5$\,ps, which corresponds to a beating period of $2T_{g_1} = 386.9$\,ps. Panel (d) shows the extracted phase difference, $\Delta\phi$, between the two orbitals as a function of time. This experimental dependence is linear, with a slope corresponding to the period of the beats. The energy splitting between two states extracted from this slope is $\vert\Delta E\vert = 10.6\,\mu\text{eV}$.}
    \label{fig:6}
\end{figure}

To determine the state of the condensate, its radiation was collected through the same objective through which the optical pumping was carried out. The incident beams and polariton emission were separated using a non-polarizing beam splitter. Laser light in the detection channel was cut off using a low-pass filter. To measure time- and wavelength-integrated photoluminescence a cooled CCD camera was used. Part of the emitted light was passed through a polarization analyzer, consisting of a quarter- and half-wave phase plate and a linear polarizer, allowing the measurement of various components of the polarization vector. The observed emission of the polariton laser was linearly polarized with the polarization degree exceeding 50 \%. The predominant polarization component was the linear horizontal polarization, see figure~\ref{fig:5}. A Mach-Zehnder interferometer was used to analyze the phase distribution of the polariton emission. Scanning the length of one of the interferometer arms allowed us to measure the coherence time $g_1(\tau)$ of the condensate (figure~\ref{fig:6}). 
\ak{This figure shows the first order coherence of the system measured by a time-resolved interferometry at different spatial locations (shown by points of different colors). The coherence time extracted from these measurements ranges from about 750 ps to about 1200 ps that is two orders of magnitude longer than the polariton lifetime in our system. Panel (d) in figure~\ref{fig:6} shows the time-dependent phase difference between two orbitals that serve as a basis of our qubit. Remarkably, the phase difference depends linearly on time, which is a signature of quantum beats in a linear system. The slope of this dependence corresponds to the energy splitting of the eigen-states of the qubit.}

\begin{figure}[b]
    \centering
    \includegraphics[width=0.9\linewidth]{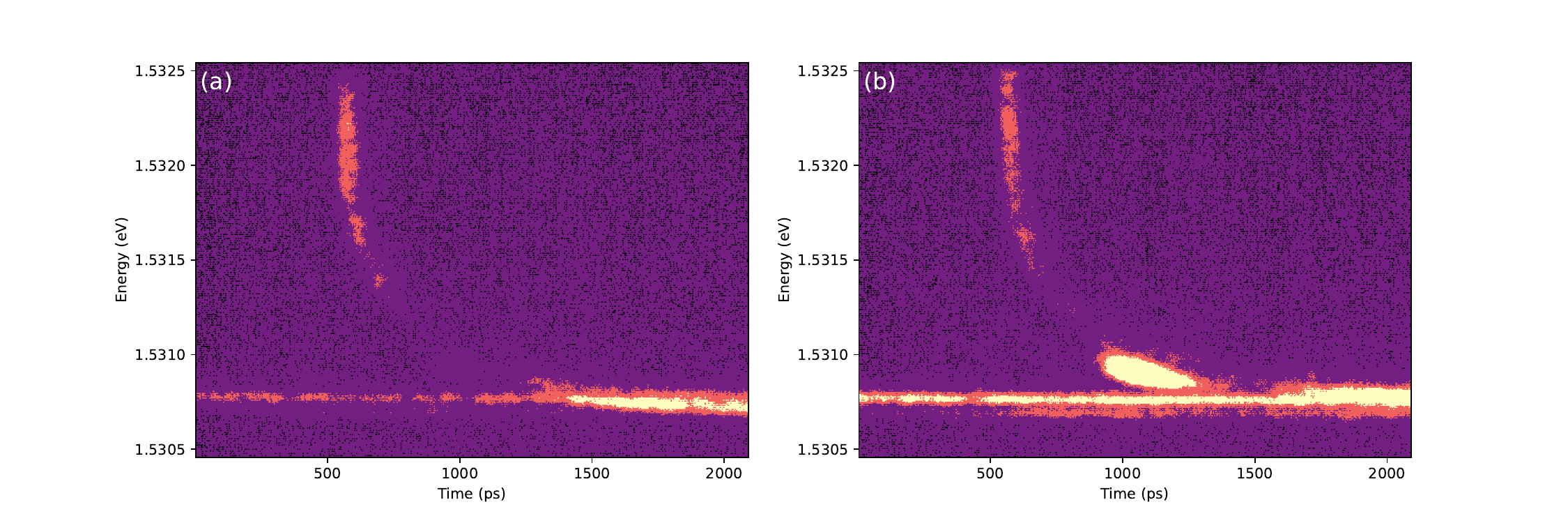}
    \caption{Time- and energy-resolved emission of the microcavity after high energy femtosecond pulse excitation arriving at $t = $600 ps. Horizontal line at 1.530 eV is the emission of condensate in the trap. (a) and (b) panels shows emission taken far and near from the pulse. }
    \label{fig:7}
\end{figure}

\newpage

\begin{figure}[t]
    \centering
    \includegraphics[width=0.90\linewidth]{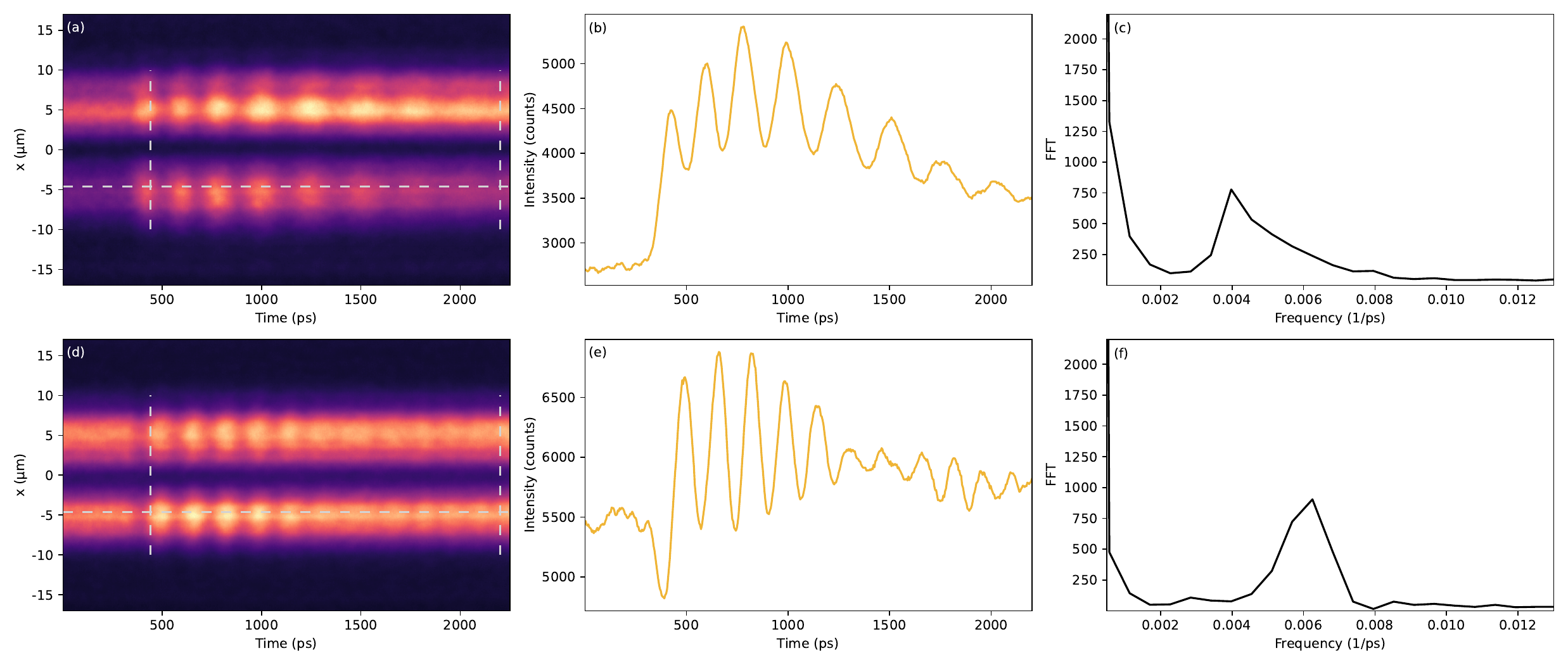}
    \caption{(a), (d) Time- and spatially-resolved emission of the polariton condensate excited with a single femtosecond pulse. Variations of shaping of the laser beam forming the trap leads to the intensity oscillations with a time-dependent frequency,  (a), oscillations with two characteristic frequencies, (d). The time dependence in panels (b), (e) corresponds to the white dashed line in panels (a), (d), respectively. Panels (c), (f) show the energy-resolved spectra calculated from the time dependent intensities (b), (e), respectively.}
    \label{fig:8}
\end{figure}

At the other output of the spectrometer, a streak camera with a synchro-scan unit was installed, providing a time resolution of 5 ps on a time scale of 2 ns. To sequentially scan the image of the condensate along the slit of the streak camera, a focusing lens in front of the spectrometer was installed on a motorized stage, setting a scanning step in terms of distance on the sample surface of down to 10 nm. The sample has a certain density of defects, which appear as vertical and horizontal lines in the radiation of the polariton condensate. Apart from affecting the polariton condensation, these defects do not manifest themselves in any way, but the condensate pinned to them is seen more brightly than if placed in the rest of the sample. The distance between the defects is on the order of 1-20 micrometers depending on the point on the sample. The dipole (dumbbell) state in the trap interacts with these defects, which manifests itself in a change in the orientation of the dipole as it approaches and moves away from the defect.

\begin{figure}[b]
    \centering
    \includegraphics[width=0.8\linewidth]{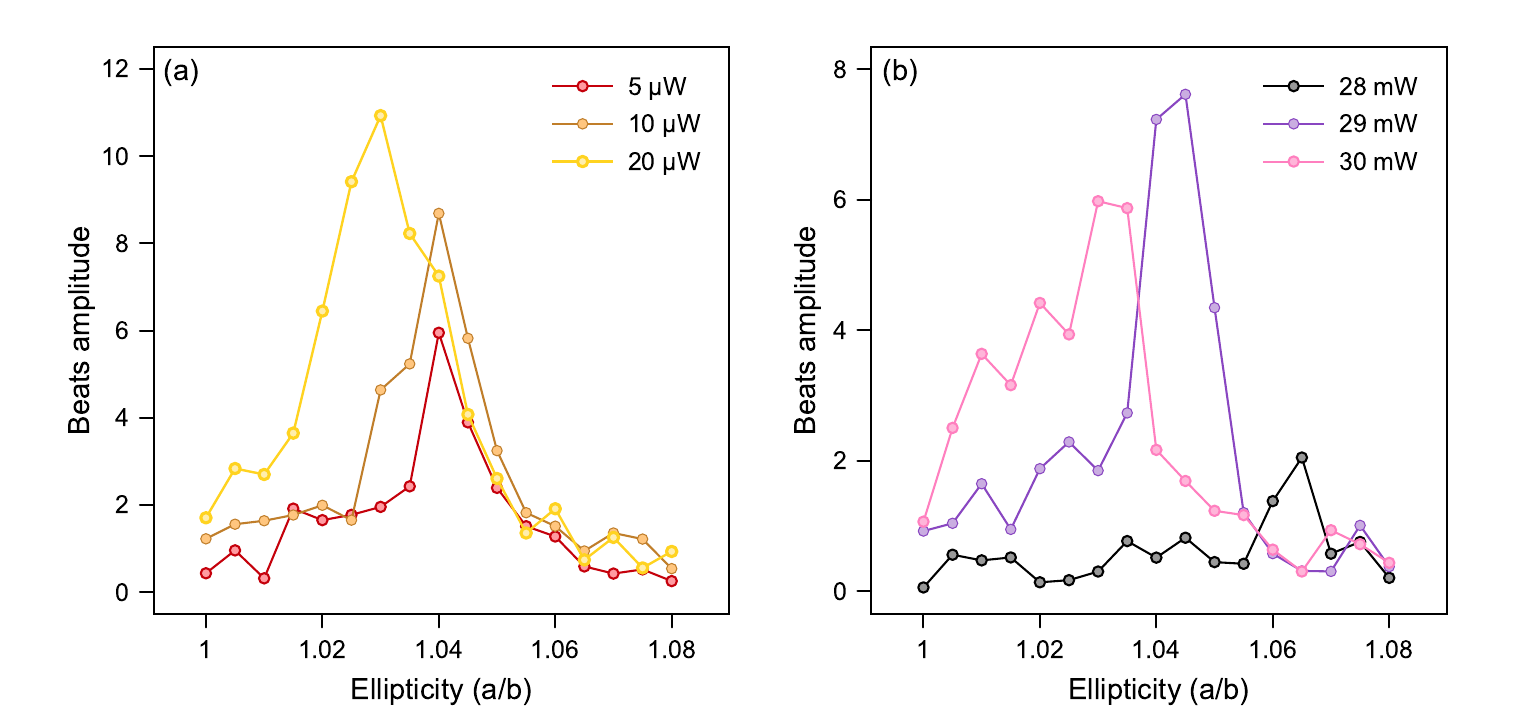}
    \caption{The influence of the pump power of the control (a) and cw (b) laser beams on the position and amplitude of the parametric resonance that occurs once the repetition rate of laser pulses matches the splitting between the excited states of the trap. The energy splitting of the eigen states of the trap is controlled by the ellipticity (a/b) of the trap as shown in figure~\ref{fig:14}.}
    \label{fig:9}
\end{figure}

\newpage

\begin{figure}[t]
    \centering
    \includegraphics[width=0.90\linewidth]{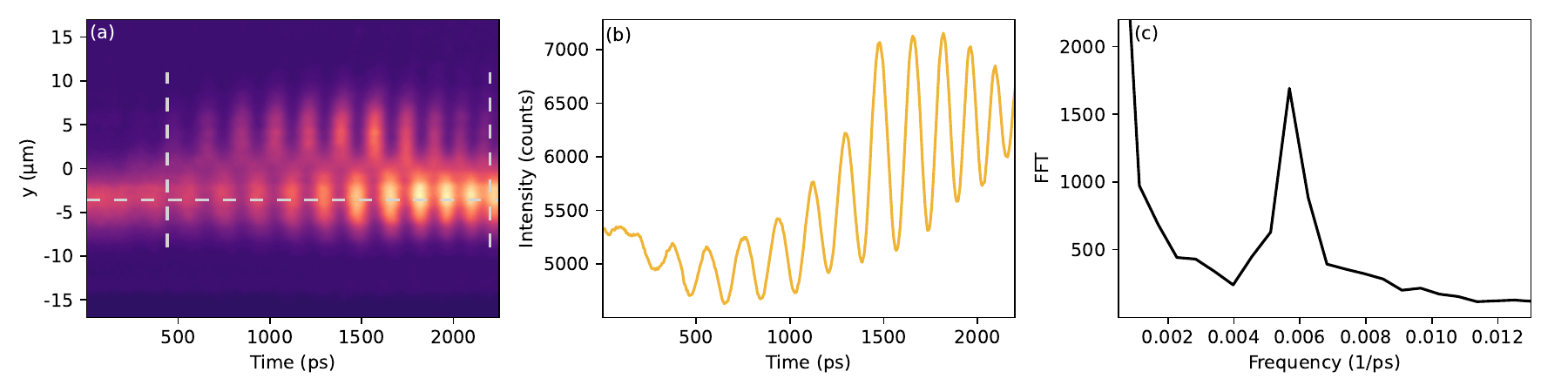}
    \caption{(a) Sequential pumping of condensate oscillations with a series of eight femtosecond pulses. The first pulse arrives at t = 200 ps. The last pulse arrives at t = 1530 ps. The time range of 1530-2200 ps corresponds to the decay of oscillations after the train of control pulses. (b) Cross-section of the time- and spatially-resolved image taken at the maximum of the condensate emission. (c) The Fourier spectrum of emission of the condensate calculated from the cross-section of the time-dependent emission in the time range between the vertical dashed lines. }
    \label{fig:10}
\end{figure}

Depending on the location of the spot on a condensate where the control pulse has arrived, different dynamics of the condensate evolution can be observed. In this work we discuss only the case where this dynamics shows periodical oscillations characterized with a large number of periods. Even in a perfectly symmetrical trap we observe oscillations. Their frequency depends on the relative positions of the trap and defects, as well as on the power of the control pulse and the power of the cw pump laser. These oscillations may be assigned to a spontaneous symmetry breaking induced by non-Hermitian terms in the Hamiltonian of the system. These terms have a little effect on the dynamics of the system once it is initialized in a superposition of two eigen-states of a trap, as our analysis shows. Indeed, the observed quantum beats, their phase and amplitude are governed by a conservative linear Hamiltonian, as our calculations show. Still, the presence of non-linear and non-Hermitian terms may play a minor role shifting the frequency of quantum beats by a small constant value. In addition, at single pulse excitation, the beats frequency usually changes over time, decreasing almost twice as time goes, see figure~\ref{fig:8}. In the spectrum of the beat signal, this manifests itself as the appearance of a long-wavelength tail. Sometimes the beats at multiple close frequencies can be also  observed, as figure~\ref{fig:8}b shows. For a control pulse train consisting of 8 pulses, no beats occur in a symmetrical trap, since the pulse repetition rate does not coincide with the random beat frequency defined by the defects. To find the ellipticity at which the splitting between states coincides with the pulse repetition rate we scan the ellipticity once the distance between pulses in the train is fixed. To do this, patterns with ellipticity ranging from 1 to 1.1 are alternately displayed on the SLM and the beat amplitude is measured. An example of such a dependence is presented in figure~\ref{fig:9}. The observed dependence of the beats amplitude on the ellipticity usually exhibits a narrow maximum, the position of which depends both on the power of the control pulse (a) and on the power of the cw laser creating the trap (b). The resulting beats are largely monochromatic, since we are selectively pumping a specific transition (figure~\ref{fig:10}.) 

\newpage
\clearpage

\begin{figure}[t]
    \centering
    \includegraphics[width=0.95\linewidth]{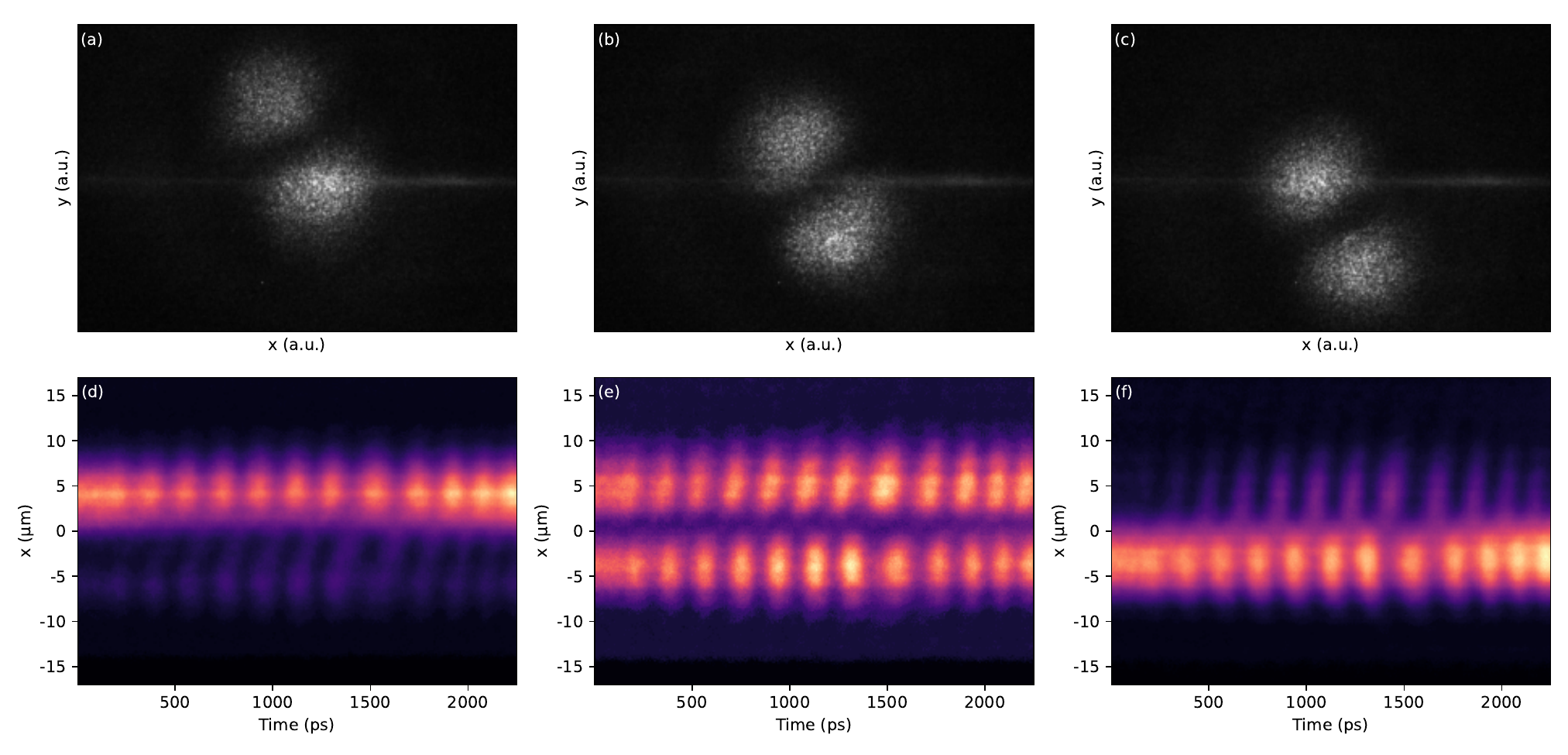}
\caption{Experimentally measured time- and specially-resolved emission of the polariton condensate at different cross-sections. (a)-(c) Relative positions of the condensate emission map in the real space and the streak camera slit (horizontal line in the middle of the panels). (d)-(f) Corresponding time dependencies of the condensate emission. The excitation scheme corresponds to the consecutive implementation of the Hadamard and Pauli-Z operation. }
    \label{fig:11}
\end{figure}

To measure the time dependence of the emission of the condensate, the latter is divided into successive strips (figure~\ref{fig:11}(a-c)). In each strip on the streak camera, the time dependence of the emission is measured (figure~\ref{fig:11}(d-f)), the the accumulated data are assembled by a computer to create a single time-dependent image. Based on this array of data, gif images were built in which the time scale was increased by $10^{10}$ times, see the Supplementary Data. The comparison with theoretical predictions was done by extracting of the angle of rotation of the condensate in real space. To do this, at each point in time, a circular cross-section of the condensate in the region of maximum luminescence was taken and the dependence of the emission intensity was plotted as a function of the polar angle assuming that the center of coordinates coincides with the center of the condensate (figure~\ref{fig:12}) This dependence is well described by the square cosine function characterised with a certain amplitude and phase. As a result of the approximation of successive sections, the dependence of the phase on time was extracted, which was then compared to the theory. 

\begin{figure}[b]
    \centering
    \includegraphics[width=0.65\linewidth]{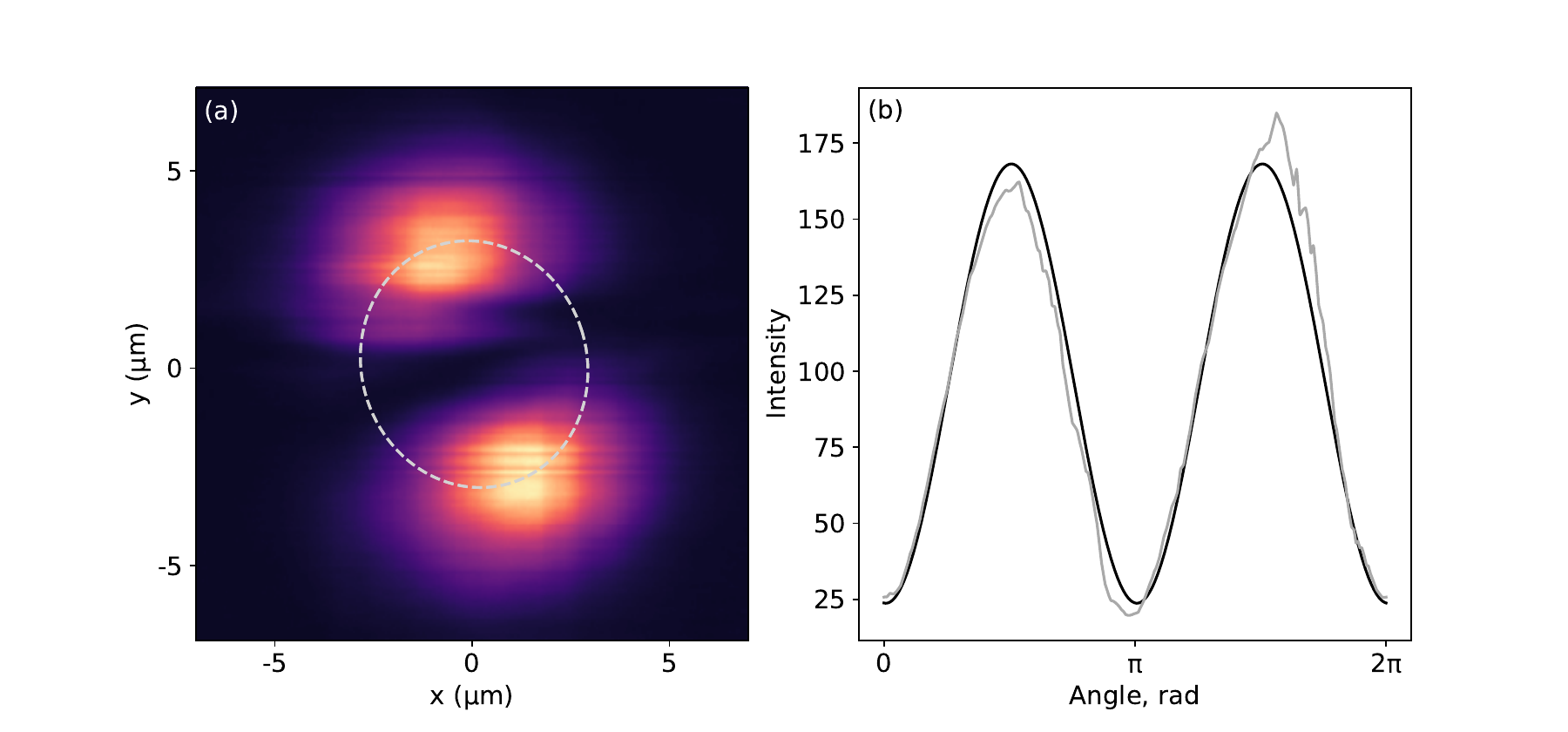}
      \caption{(a) An example of the extraction of the tilting angle of the dipole (dumbbell) state from the reconstructed image of the condensate. The yellow dashed line shows an angular section where the condensate intensity is sampled for the approximation. (b) The yellow curve shows the emission intensity as a function of angle around the center of the image. The fit of the date with a function $y(\phi)=Acos^2(\phi + \alpha)$, where $\alpha$ is the tilting angle of the dipole, is shown by the black line. }
    \label{fig:12}
\end{figure}

\newpage
\clearpage

\begin{figure}
    \centering
    \includegraphics[width=0.8\linewidth]{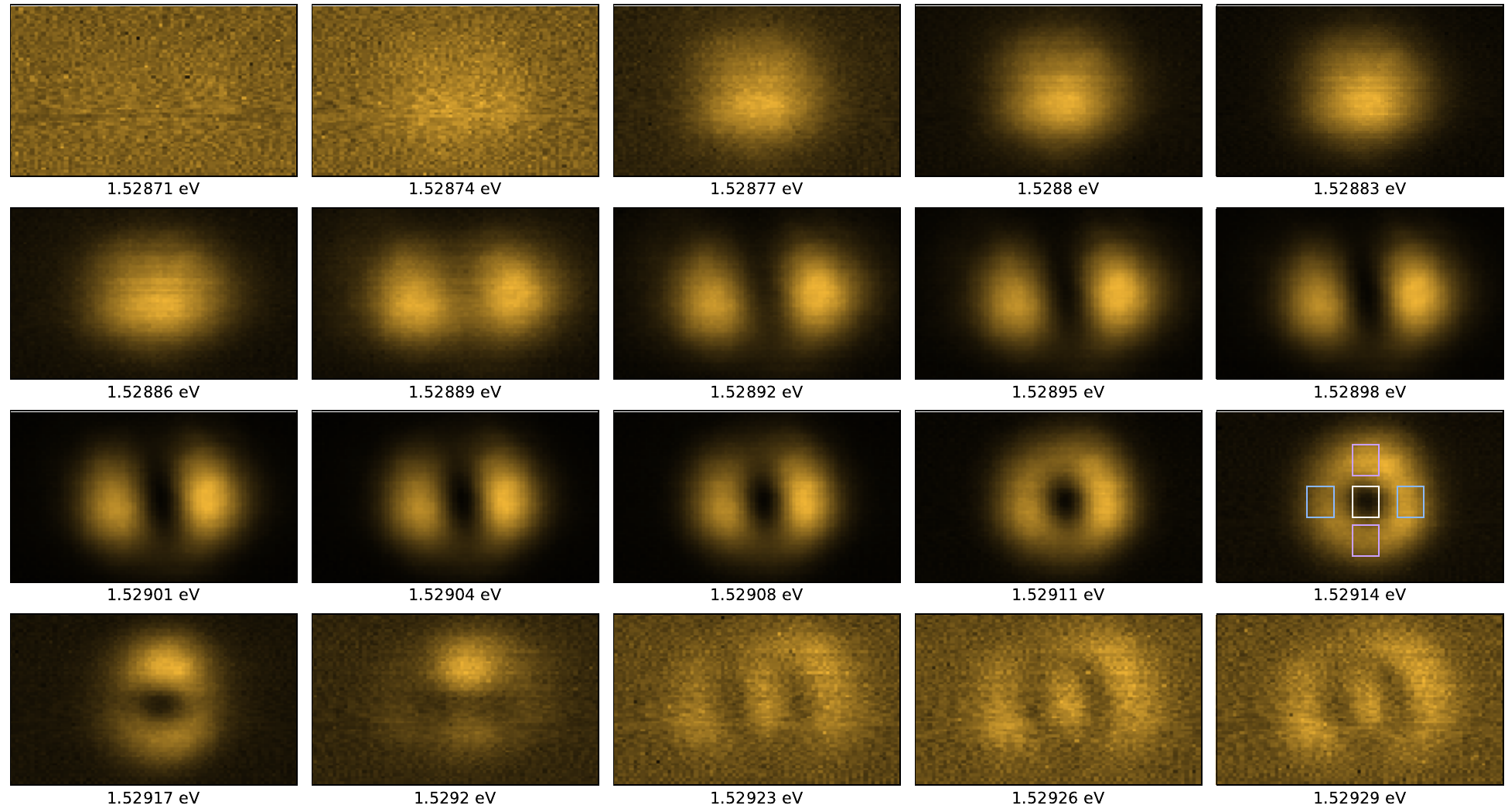}
    \caption{Spatial distribution of the energy-resolved emission of the polariton condensate confined in an elliptical trap. Sequential panels correspond to the sequential pixels of the CCD camera placed at the output slit of the spectrometer. Each image is normalized to the maximum intensity. At the highest intensity of the cw laser that forms the trap all the excited states of the trap are populated by polaritons. }
    \label{fig:13}
\end{figure}

 The splitting between dipole (dumbbell) states can be extracted from the frequency of the beats and it can also be directly measured spectroscopically. For this purpose, the condensate radiation has been also decomposed into successive strips, but instead of measuring the time dependence, this time we have measured the frequency-resolved spectrum of emission in each strip. Then, using a computer, an image of the condensate was assembled and plotted as a function of the emission energy as figure~\ref{fig:13} shows. In this case, instead of control pulses, we used a higher pump power in such a way that all states confined in the trap were populated. After processing, only two dipole-like states were selected from the full set of states, and their spectral position was determined as a function of the trap ellipticity. Although the experiment was performed at a different pump power, a good agreement was found between the splitting frequency extracted from the beat frequency and one obtained by the spectral analysis (figure~\ref{fig:14}).
 
\begin{figure}[b]
    \centering
    \includegraphics[width=0.53\linewidth]{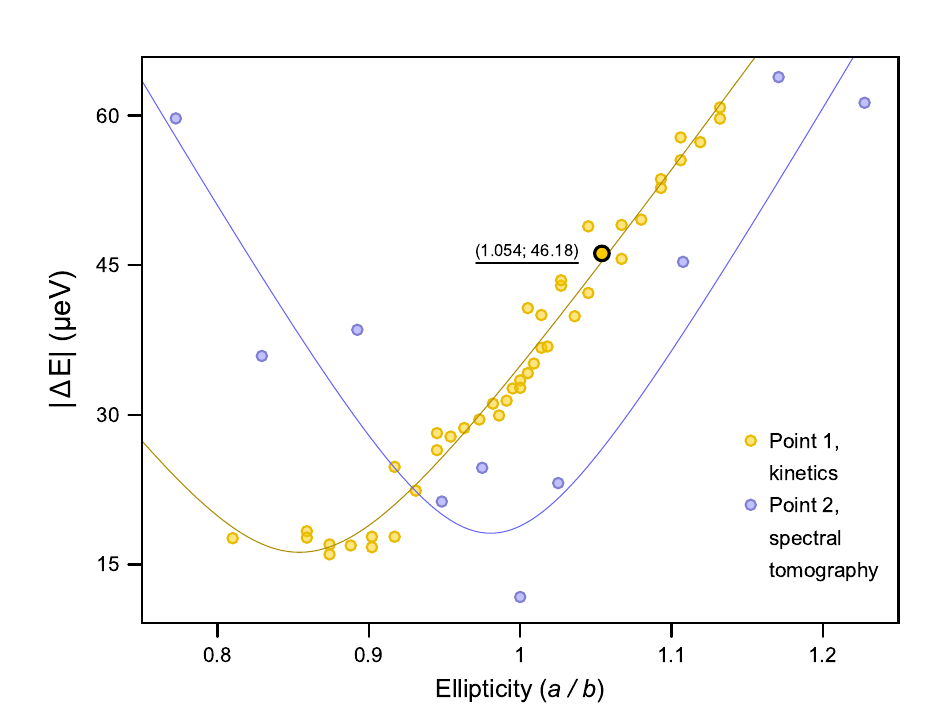}
    \caption{Energy splitting between the 2\textsuperscript{nd} and 3\textsuperscript{rd} dipole states in the trap as a function of the ellipticity of the trap extracted from the frequency of intensity oscillations (yellow) and the spectroscopy data (lilac). The curves are theoretical fits. }
    \label{fig:14}
\end{figure}

\newpage
\clearpage

\section*{Model}

A polariton condensate in an elliptical trap is characterised by a discrete energy spectrum. The eigen energies are not equidistant, and splittings between them may be efficiently tuned by controlling the ellipticity of the trap. In the limit of small ellipticity, the splitting between 2nd and 3rd energy levels corresponding to $p_x$ and $p_y$ orbitals is much smaller than the splittings between 1st and 2nd or 3rd and 4th energy levels. This allows to approximate the coherent dynamics of a superposition of $p_x$ and $p_y$ states by a 2x2 Hamiltonian. The beats observed in our experiments can be reproduced by a linear two-level model with a good accuracy. This allows us to consider $p_x$ and $p_y$ states of the trapped condensate as the computational basis states $|0\rangle$ and $|1\rangle$. 

In order to make our model quasi-analytical, let us describe the trap confining the condensate of polaritons using the potential of a two-dimensional harmonic oscillator
\begin{align}\label{eqn:pot_basic}
    V=\dfrac{m f(\varphi)}{2}\left(\omega_x^2 x^2+\omega_y^2 y^2\right),
\end{align}
where the difference between $\omega_x$ and $\omega_y$ accounts for the  controlled ellipticity.
We introduce a dimensionless function $f(\varphi)$ (of the order of $\sim 1$) that accounts for the built-in anisotropy of the trap. The potential~\eqref{eqn:pot_basic} enters the Schr\"{o}dinger equation
\begin{align}\label{eqn:schrod_eq_sm}
    &i\hbar\pdv{\psi}{t} = \left(-\frac{\hbar^2}{2m}\Delta + V\right)\psi = \hat H_0\psi.
\end{align}
The stationary solutions of such a Hamiltonian can be found numerically. In the case of a small ellipticity ($a=\omega_x/\omega_y\approx 1$) the first two excited states are $\psi_{10}$ and $\psi_{01}$, which differ in energy by $\Delta E = \hbar(\omega_x - \omega_y)$.
Numerically, we obtain a similar energy spectrum of the trap. Let us denote $p_x$ and $p_y$ orbitals as $\ket{0}$ and $\ket{1}$ and their energies as $\varepsilon_0$ and $\varepsilon_1$, respectively. To describe the beats in the condensate, we set the initial state of the system as follows:
\begin{equation}
    \ket{\psi} = \cos(\frac\theta2)\ket{0} + \sin(\frac{\theta}{2})e^{i\phi}\ket{1},
\end{equation}
where $\theta$ and $\phi$ are the latitude and the longitude on the Bloch sphere, respectively.

The evolution of the state $\ket\psi$ as a function of time $\tau$ in the basis of the eigen-functions of a non-perturbed system $\ket 0$ and $\ket 1$  can be described by the following operator:
\begin{align}
    \hat U_0(\tau) = \exp(-\frac{i}{\hbar}\hat H_0\tau) = 
     \mqty[\exp(-i\varepsilon_0 \tau/\hbar) & 0 \\ 0 & \exp(-i\varepsilon_1\tau/\hbar)].
\end{align}

To change the state of the condensate, or to put it into the oscillatory mode it is necessary to employ a control pulse. We assume here that is initially the system finds itself in one of the dipole states $\ket{0}$ or $\ket{1}$).  We describe the impact of the pulse with use of the perturbation theory. Namely, we introduce the perturbation potential $V_{\textup{I}}(\vb{r}, t) = V_{0,\textup{I}} f(\vb{r})g(t)$, which is localised in the real space near the point $\vb{r}_0=(x_0,\, y_0)$ : $f(x,y)=\exp[-\alpha(\vb{r}-\vb{r}_0)^2]$ and and in time near the moment $t_0$: $g(t)=\exp[-\gamma(t)(t-t_0)^2]$, where $\gamma(t)=\gamma_1$ for $t<t_0$, and $\gamma(t)=\gamma_2$ for $t>t_0$.

The Hamiltonian of a perturbed system reads $\hat H = \hat H_0 + V_{\textup{I}}(\vb{r},t)$. The matrix elements $v_{ij}(t)$ of the pulse in the basis of $\ket{0}$ and $\ket{1}$ at the moment $t$ can be calculated numerically.
The evolution of $ \ket{\psi}$ is described now by the following operator 
\begin{equation}
    \hat U_I(t_1, t_2) = \mathcal{T}\qty{\exp\left[-\frac{i}{\hbar}\int_{t_1}^{t_2}\hat H(t')\dd t'\right]}, \quad     \hat{H}=\mqty[\varepsilon_0 + v_{00}(t) & v_{01}(t) \\ v_{10}(t) & \varepsilon_1 + v_{11}(t)].
\end{equation}

We assume the perturbation potential to be small enough in order to enable one to neglect the mixing of $p_x$ and $p_y$ states with the ground state or higher excited states. 

\newpage
\clearpage

\subsection*{Simulation of a quantum gate operation: the Hadamard transform}

Let us consider system being in the quantum state $\ket{\psi_0}$ at the moment $t=0$. The action of a Hadamard gate at this state should bring the system to the state $\ket{\psi_H}$ at the time $\tau > 0$:
\begin{equation}
    \ket{\psi_H} = \hat U_0(\tau)\hat H\ket{\psi_0},
\end{equation}
where $H$ is defined as:
\begin{align}\label{eq:hadamar_gate_matrix_2}
H = \frac{1}{\sqrt{2}} \begin{bmatrix} 1 & 1 \\ 1 & -1 \end{bmatrix}.
\end{align}
In practice, the Hadamard gate can be implemented by applying a time-dependent perturbation potential   $V_{\textup{I}}({\bf r},t)$. The action of this potential at times $t_0 > 0$ transfers the initial state $\ket{\psi_0}$ to the state $\ket{\psi_I}$ at the time $\tau > 0$:
\begin{equation}
    \ket{\psi_I} = \hat U_I(0, \tau)\ket{\psi_0}.
\end{equation}
Let us consider $\tau > t_0$ long enough for the potential to extinct:
\begin{equation}
    \exp[-\gamma_2(t_0-\tau)^2]\ll 1
\end{equation}
In this limit, the evolution of state $\ket{\psi_I}$ is the same as one in a non-perturbed system. Thus, the chosen pulse triggers the Hadamard gate operation if the states $\ket{\psi_I}$ and $\ket{\psi_H}$ differ only by the phase $e^{i\varphi}$ for any initial state $\ket{\psi_0}$. In other words, the following condition must be satisfied
\begin{equation} 
    \hat U_I(0, \tau) = e^{i\varphi}\hat U_0(\tau) H.
\label{evolution_operators_equiv}
\end{equation}

\begin{figure}[b]
    \centering
    \includegraphics[width = 0.45\linewidth]{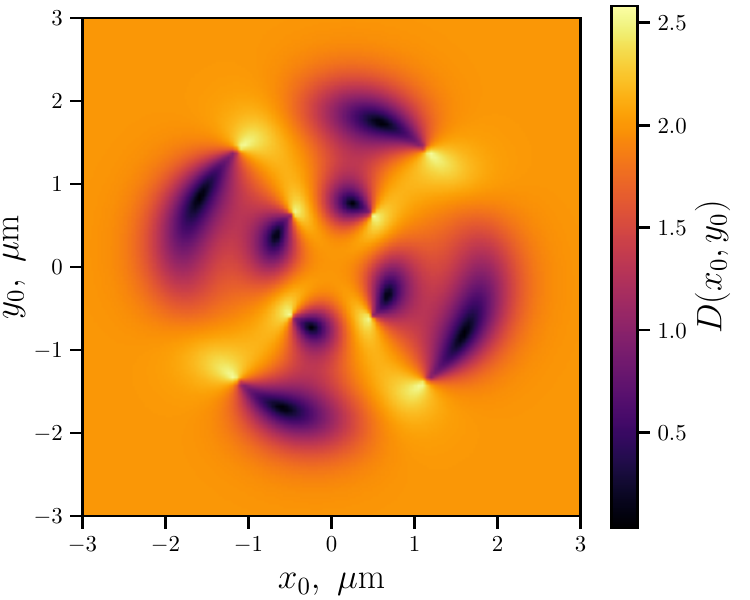}    \includegraphics[width=0.42\linewidth]{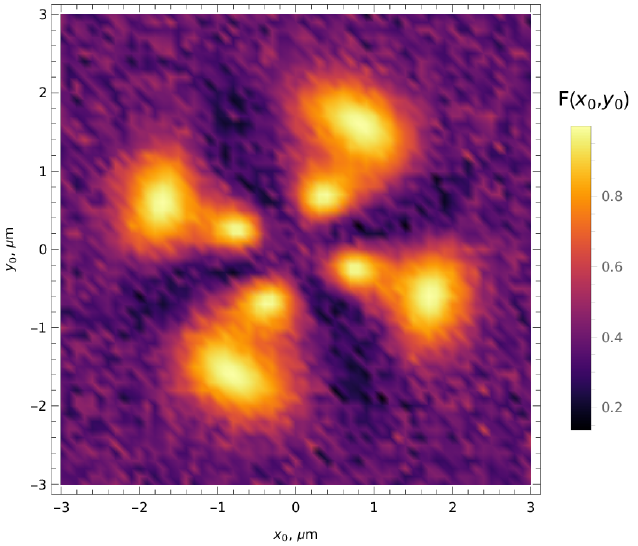}
    \caption{\label{fig:F_x0y0}(left panel) The dependence of the magnitude D at the center of the pulse $(x_0, y_0)$ for the Hadamard operator at $t_0=292.2$ ps. (right panel) Dependence of the average fidelity $F$ (found via averaging over 30 randomly selected $\psi_0$) on the position of the center of the control pulse spot $(x_0, y_0)$ at $t_0=292.2$ ps. This dependence should coincide with those of figure~\ref{fig:D_x0y0}, in the limit of infinite number of considered initial states.\label{fig:D_x0y0}}
\end{figure}

\newpage

\subsection*{The optimization of parameters of the perturbation potential}

\begin{figure}[b]
    \centering
    \includegraphics[width=0.5\linewidth]{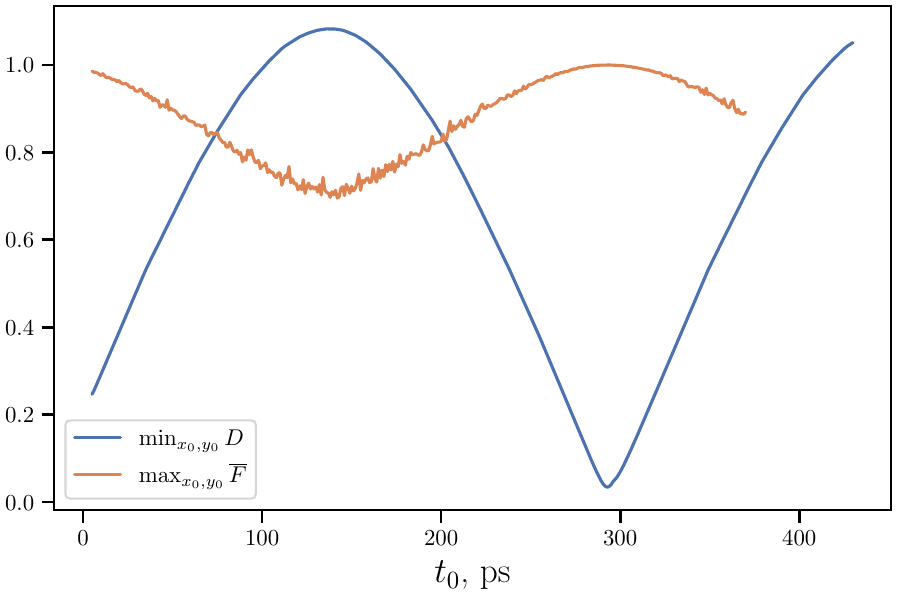}
    \caption{The dependence of the longest distance $D$ (by $x_0$, $y_0$) and the highest fidelity of the Hadamard operation F (by $x_0$, $y_0$) found on a set of calculated evolutions of 30 randomly selected initial states $\psi_0$ on the arrival time of the control pulse $t_0$ .}
    \label{fig:D_F_t0}
\end{figure}

To obtain the control pulse parameters needed for the most accurate implementation of the Hadamard gate we should maximize a fidelity $F$, which is canonically defined through the dot product of the anticipated (true) and actually obtained Bloch vectors averaged over a large number of different initial states:
\begin{equation}
    F \equiv \bar{F} = \langle\abs{\braket{\psi_H}{\psi_I}}^2\rangle.
\end{equation}
$F=1$ would indicate that the chosen perturbation pulse triggers the Hadamard operation with a hundred percent fidelity.

There is another way to find the required parameters of the perturbation potential, which is less demanding from the computational point of view. To obtain the evolution operator $\hat U_I(0, \tau)$ of the system perturbed by $V_{I}({\bf r},t)$, we can use the fact that the corresponding columns of evolution operator matrix describe the evolution of states $\ket 0$ and $\ket 1$ which are vectors $(1,\, 0)$ and $(0,\, 1)$, respectively, in the basis $\psi_{10}$ and $\psi_{01}$ from $t=0$ to $t=\tau$. The parameters of the pulse satisfying the equation (\ref{evolution_operators_equiv}) can be found by minimization of $L_2$ norm of the difference between $\hat U_I(0, \tau)$ and $ \hat U_0(\tau) H$ taking into account the possible difference in phase:

Below we give an example of calculation for the following set of parameters: $m = 510$ meV, $\omega_x = 0.2$ ps$^{-1}$, $\omega_y = 0.21$ ps$^{-1}$, $V_{\textup{I},0} = 0.4$ meV, $\gamma_1 = 5 \cdot 10^{-2}$ ps$^{-2}$, $\gamma_2 = 5 \cdot 10^{-4}$ ps$^{-2}$. The dependencies of the matrix distance $D$ and the average fidelity $F$ on $x_0$ and $y_0$ are shown in figure~\ref{fig:D_x0y0}. Figure~\ref{fig:D_F_t0} shows the lowest (by $x_0,y_0$) $D$ and the highest (by $x_0,y_0$) $\overline{F}$ dependencies on the arrival time of the control pulse $t_0$. These dependencies demonstrate that the higher fidelity $\overline{F}$ corresponds to the lower norm $D$, moreover, the introduced metric $D$ yields more reliable estimates of the correct time of control pulse arrival. With this procedure we obtain an average fidelity of $\sim 0.9999$ for $30$ randomly selected $\ket{\psi_0}$ at the optimised pulse location (introduced below) at $t_0=292.2$ ps.

\begin{align}
    D = \norm{\hat U_I(0, \tau)e^{i\xi} - \hat U_0(\tau) H}, \quad 
    \xi\!\!=\! \arg\qty[\qty(\hat U_0(\tau) H)_{00}] - \arg\qty[\qty(\hat U_I(0, \tau))_{00}].
\end{align}

\newpage
\clearpage

One can see that $\ket{\psi_H}$ and $\ket{\psi_I}$ are almost the same states and the pulse simulates Hadamard gate with a good accuracy. The lowest norm $D$ for the Hadamard operator has been obtained for $x_0 = -1.72$ $\mu$m, $y_0 = 0.59$ $\mu$m and at $t_0 = 292.2$ ps (the maximal fidelity is found for the same coordinate). In figure~\ref{fig:BS_exmp}, we also demonstrate an example of the action of the perturbation potential in the case of obtained coordinate and sending time. As an agreement, one can choose that such a point in time should be determined by one percent of the maximal value of the control pulse, in case of $t_0=292.2$ ps, this moment of time is $282.5$ ps.

\begin{figure}[t]
    \centering
    \includegraphics[width=0.25 \linewidth]{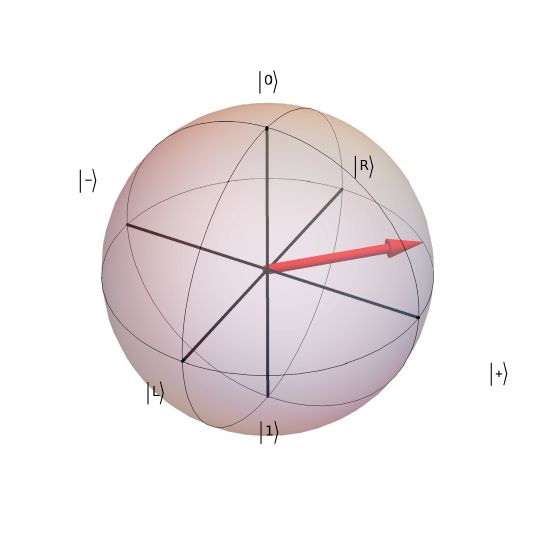}
    \includegraphics[width=0.25 \linewidth]{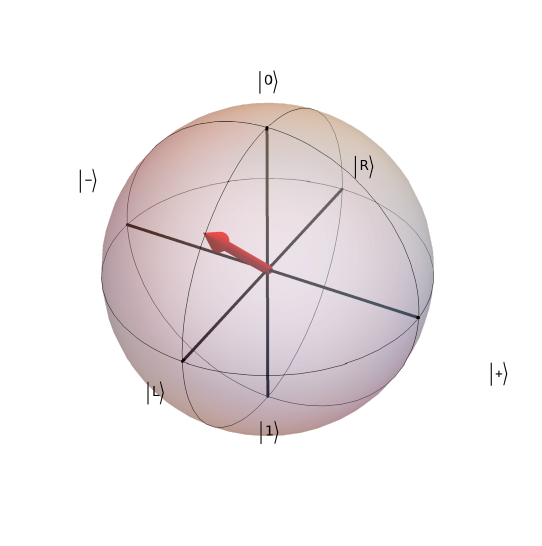}\\
    \includegraphics[width=0.25 \linewidth]{smpic_th//BlochSpherePsi0.pdf}
    \includegraphics[width=0.25 \linewidth]{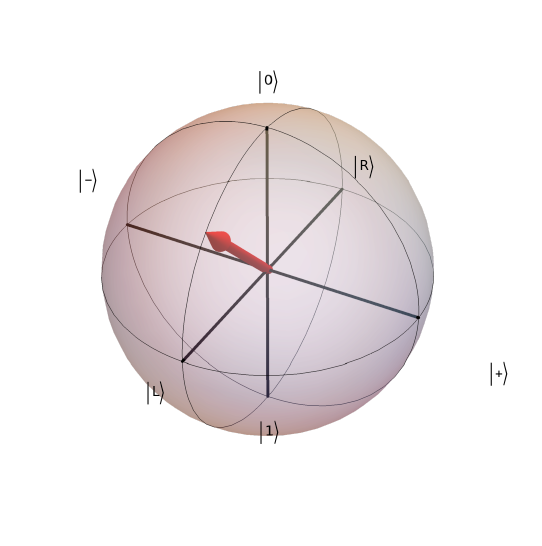}
    \caption{Initial and final states of the system perturbed by the control pulse compared to the anticipated result of the Hadamard operation visualized on the Bloch sphere. Upper panel shows the initial state $\ket{\psi_0}$ (left) and the final state $\ket{\psi_I}$ after the pulse action (right). The lower panel shows the same initial state $\ket{\psi_0}$ (left) and the target state $\ket{\psi_H}$ corresponding to the Hadamard gate operation (right). The final state is detected at the time $\tau = 500$ ps.}
    \label{fig:BS_exmp}
\end{figure}

\begin{figure}[b!]
    \centering
    \includegraphics[width = 0.49\linewidth]{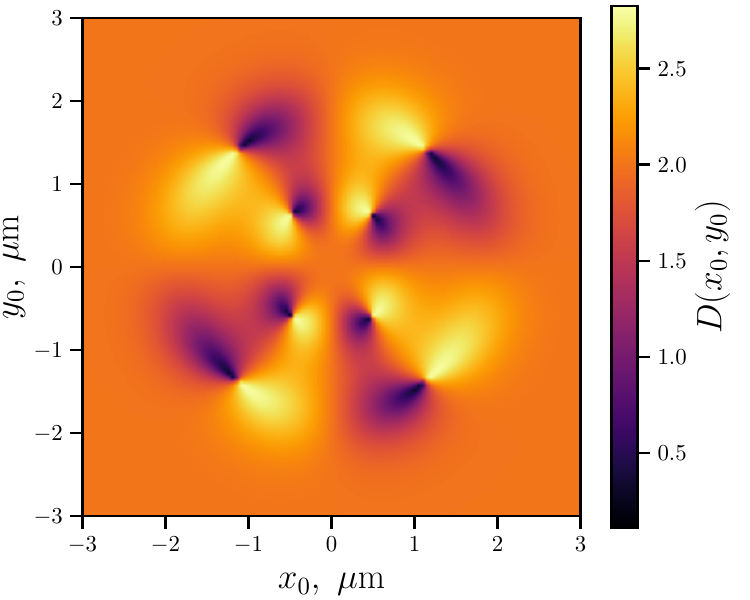}\includegraphics[width = 0.49\linewidth]{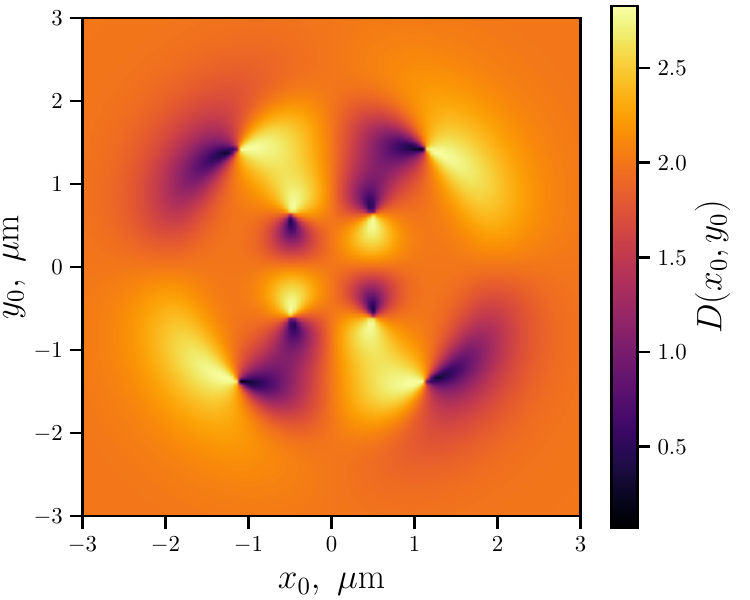}
    \caption{The dependencies of the norm $D$ on the location of the center of the control pulse spot $(x_0, y_0)$ for $\sigma_x$(left panel) operator at $t_0=292.2$ ps and $\sigma_y$ (right panel) operator at $t_0=135$ ps.}\label{fig:sigma_xy_heat_map}
\end{figure}

\begin{figure}[t!]
    \centering
    \includegraphics[width = 0.55\linewidth]{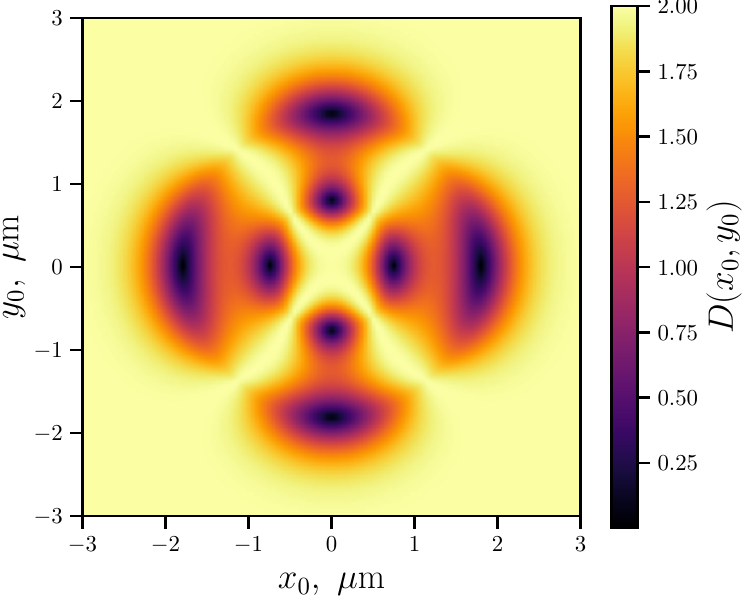}
    \caption{The dependence of the norm $D$ on the location of the center of the pulse spot $(x_0, y_0)$ for $\sigma_z$ operator at $t_0=292.2$ ps.}\label{fig:sigma_z_heat_map}
\end{figure}

\begin{figure}[t!]
    \centering    \includegraphics[width=0.7\linewidth]{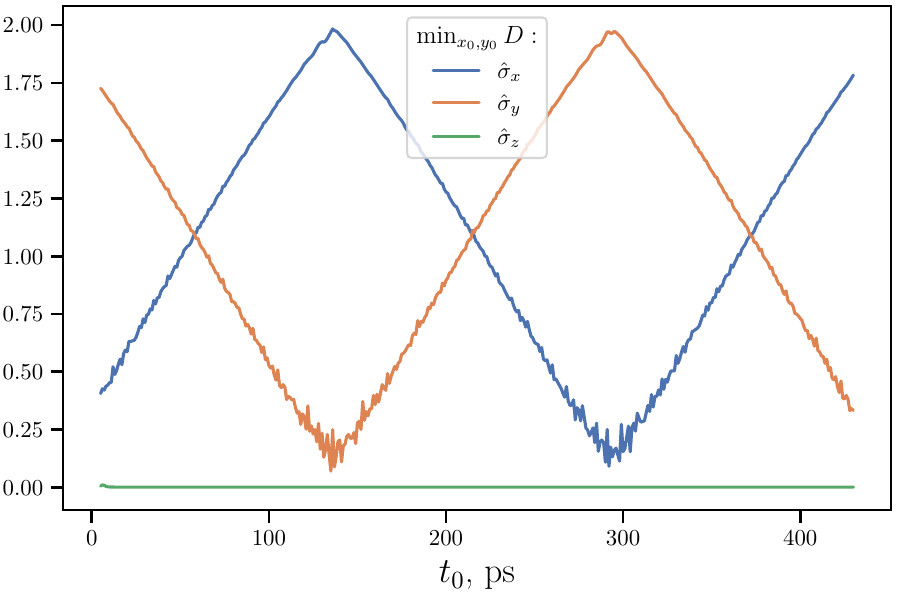}
    \caption{The dependence of the shortest distance $D$ (by $x_0$, $y_0$) on time of the control pulse arrival $t_0$ for all Pauli operationss.}
    \label{fig:xyz_norms}
\end{figure}

\newpage
\clearpage

\subsection*{The realization of Pauli operations}

The similar maps can be obtained the operators which correspond to Puali matrices: $\sigma_x$, $\sigma_y$ and $\sigma_z$. The lowest norm $D$ for $\sigma_x$ operator has been obtained at $x_0 = 0.6$ $\mu$m, $y_0 = -0.5$ $\mu$m and $t_0 = 292.2$ ps, the corresponding heat map is presented in figure~\ref{fig:sigma_xy_heat_map}. The lowest norm $D$ for $\sigma_y$ operator has been obtained for $x_0 = 1.4$ $\mu$m, $y_0 = -1.1$ $\mu$m and $t_0 = 135$ ps.  The corresponding color map is presented in figure~\ref{fig:sigma_xy_heat_map}. Finally, the lowest norm $D$ for $\sigma_z$ operator has been obtained for $x_0 = 0$ $\mu$m, $y_0 = 1.8$ $\mu$m and $t_0 = 66$ ps. The corresponding color map is shown in figure~\ref{fig:sigma_z_heat_map}. We note that the norm distribution for $\sigma_z$ is independent on time, this gate can be realized at any time moment by sending the control pulse to the specific spots. Finally, we summarize all the data on the time dependencies of the best-norm behaviour on time in figure~\ref{fig:xyz_norms}.

\subsection*{The theoretical description of the decay of quantum beats}

The experimental data demonstrates a damping of oscillations that leads to the eventual relaxation of the condensate to $\ket{0}$ state. In order to describe this decay and relaxation, the conservative two-level model presented in the main text needs to be upgraded. Here we account for both processes by introducing of a phenomenological exponential factor in the coefficient $\ket{1}$ of $\ket{\psi}$ superposition expansion ($\ket{\psi} = c_0(t)\ket{0} + c_1(t)\ket{1}$):
\begin{equation}
    \ket{\tilde \psi} = \frac{1}{A}\qty[c_0(t)\ket{0} + c_1(t)e^{-\beta t}\ket{1}], 
\end{equation}
where $\beta$ is a positive fitting parameter, while $A(t)=\sqrt{\abs{c_0(t)}^2 + \abs{c_1(t)}^2e^{-2\beta t}}$ is the normalization factor. An example of the application of this phenomenological approach is presented in figure~\ref{fig:angle_losses}, where behaviour of intensity corresponding to the experiment from figure~\ref{fig:3} was theoretically fitted.  

\begin{figure}[b]
    \centering
    \includegraphics[width=0.7\linewidth]{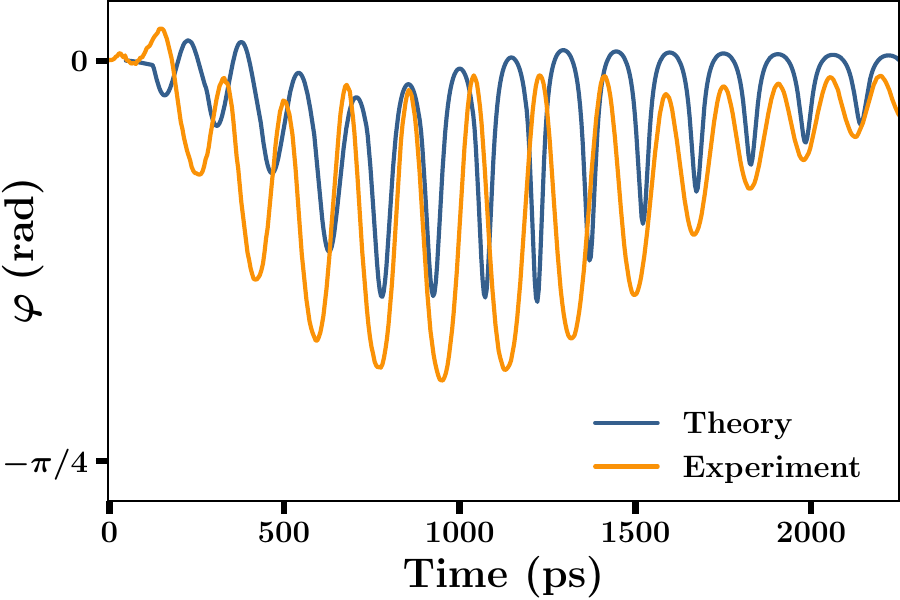}
    \caption{The time dependence of the angle between the axis of the condensate and its initial orientation: experiment and theory. This plot is a counterpart of the dynamics shown in figure~\ref{fig:3}, with damping taken into account and the parameters of the control pulse spot modified to obtain the best fit.}
    \label{fig:angle_losses}
\end{figure}

\newpage
\clearpage

\ak{

\section*{Polariton Condensates: Classical or Quantum?}

In their recent publication \cite{Barrat2024}, Barrat \textit{et al.} present a ``qubit analog'' based on a bosonic condensate of exciton-polaritons confined in a trap. They experimentally demonstrate coherent oscillations of the many-body wave function of the condensate that persist over much longer times than the individual polariton lifetime. Mapping the system onto a Bloch sphere, Barrat \textit{et al.} further demonstrate logic operations by perturbing the trap potential with non-resonant laser pulses. The ultrafast operation speed, full optical control and read-out, as well as their remarkable scalability, make trapped polariton condensates promising candidates for quantum computing \cite{Kavokin2022}.

However, before discussing applications of polariton condensates as qubits, it is crucial to ensure that they truly represent quantum objects. On one hand, the possibility of quantum computing with mesoscopic superposition states has been discussed since the beginning of the 21st century \cite{deOliveira2000}, with superconducting qubits serving as a good example of such states \cite{Kiktenko2015}. On the other hand, coherent oscillations in a polariton system—similar to those observed in \cite{Barrat2024}—may also be caused by nonlinear phenomena in classical optics, such as limit cycle dynamics \cite{CarraroHaddad2024,PhysRevLett.129.155301}. Previous observations of the oscillatory behavior of trapped polariton condensates have been given classical interpretations and described by semi-classical Gross-Pitaevskii equations \cite{PhysRevB.102.195428,PhysRevLett.126.075301}.

Can one experimentally distinguish a classical superposition state of exciton-polaritons, akin to a multimode laser, from a genuinely quantum superposition such as that described in the recent theoretical work \cite{PhysRevResearch.3.013099}? In this commentary, we formulate two criteria for the quantum nature of exciton-polariton condensates that can be verified experimentally:

\begin{enumerate}
  \item \textbf{Suppression of zero-delay intensity-intensity correlations:} If the condensate is a quantum object, a measurement may find it in either of its two basis states—but not in both simultaneously. In contrast, if the condensate is classical, a single-shot measurement of its projections onto the two basis states would yield a nonzero coincidence rate.
  
  \item \textbf{Non-zero negativity of the density matrix:} In the computational basis, the elements of the density matrix of a pair of bosonic condensates confined in tunnel-coupled traps can be measured by an optical interferometry technique similar to that in \cite{Barrat2024}. Simple algebra then allows one to deduce the negativity \cite{Vidal2002} from the density matrix elements and verify whether it deviates from zero in states that are presumably entangled. A negativity value close to 0.5 would provide strong evidence for the formation of an entangled state, akin to a Bell state.
\end{enumerate}

\section*{Classical vs. Quantum Rabi Oscillations}
Rabi oscillations in a two-level system are often considered a hallmark of quantum coherence, arising from the coherent superposition of two eigenstates. In a true quantum two-level system -- such as a single quantum dot coupled to a cavity -- the oscillations reflect the probabilistic evolution of the system's state and can manifest nonclassical features such as entanglement and number squeezing. In this regime, the evolution is governed by the linear Schr\"{o}dinger equation, and the oscillation frequency is determined solely by the energy splitting between the two states.

In contrast, classical (or semiclassical) Rabi-like oscillations can emerge in systems where a macroscopic coherent field, described by classical Bloch equations, undergoes normal-mode coupling. Although the mathematics describing the oscillations may be similar, the underlying physics is fundamentally different: in a classical oscillator, the observed dynamics result from the interference of coherent fields without genuine quantum superposition of individual particles~\cite{PhysRevLett.113.226401,Faust2013,10.1119/1.4878621}.

Our experiments probe a regime where the exciton-polariton condensate -- despite involving thousands of particles -- behaves as a single quantum entity. The observed quantum beats between the discrete $p_x$ and $p_y$ trap states are accurately modeled by a linear two-level Hamiltonian. This indicates that the system operates in a quantum regime rather than simply exhibiting classical normal-mode coupling. Such a regime is promising for implementing quantum gate operations, as evidenced by the Bloch sphere mapping and the realization of Hadamard and Pauli-$Z$ operations. The system is suitable for the implementation of quantum gate operations as confirmed by the fidelity tests.}

\newpage
\clearpage

\section*{Polariton vs superconducting qubits}

The experimental demonstration of long lasting quantum beats of a trapped polariton condensate paves the way to applications of similar systems for the realisation of quantum processors. Conceptually, polariton qubits are somewhat similar to superconducting phase qubits. In both cases, the object that is being quantized is a many body entity including millions of Cooper pairs in the case of a superconducting qubit and tens of thousands polaritons in the case of a polariton qubit. In both cases, the number of particles participating in the quantized state is linked to the phase by an uncertainty principle and cannot be known exactly. On a microscopic level, Cooper pairs constantly dissociate to uncorrelated electron pairs and are constantly being formed from uncorrelated electrons, the process that is accounted for e.g. in the BCS wave-function (see e.g. an insightful discussion in Ref. \cite{Snoke2024-ah}). Similarly, exciton-polariton polariton condensates are constantly dissipated by radiative decay and constantly replenished by stimulated scattering from excitonic reservoirs. Importantly, the spatial coherence in a superconducting circuit or in a polariton condensate seem to be not affected by fluctuations of the particle numbers caused by above mentioned processes. At least, the spatial coherence is preserved over macroscopic times that are orders of magnitude longer than the life-time of an individual quasiparticle participating in the superconducting current or a polariton condensate. 
Besides similarities, there exist important differences between superconducting qubits and (a theoretical concept of) polariton qubits. First, Cooper pairs are charged particles whose transport is strongly affected by external magnetic fields. The quantization of magnetic flux defines the energy spectra of superconductinig qubits such as fluxonium. In contrast, exciton-polaritons are electrically neutral, their quantization in a trap is nothing but size quantization. Second, Josephson junctions are paramount for the realization of superconducting qubits, as they introduce a non-linearity in dependence of the inductance of the superconducting circuit on the frequency of superconducting current. This non-linearity breaks makes the energy spectrum of a superconducting circuit non-equidistant, that allows for selection of a pair of energy split eigen states for the realization of a qubit. In contrast, the energy spectrum of a polariton condensate in a two-dimensional trap is non-equidistant anyway. In a circular trap some of the energies of some of the eigen states, such as $2p_x$ and $2p_y$ states are degenerate. This degeneracy can be lifted in an elliptical trap, where tuning the ellipticity one can efficiently tune the splitting of $2p_x$ and $2p_y$ states. This splitting remains significantly smaller than the energy separating this pair of states from $1s$ or $3s$ neighbouring states in the trap. Third, in order to control superconducting qubits, microwave pulses are used which requires fabrication of two-dimensional wave-guides. We have demonstrated that the control of polariton qubits may be achieved by using non-resonant femtosecond optical pulses. This seems a surprising finding having in mind that the optical frequency exceeds the splitting of basis states of the qubit by orders of magnitude. Still, the optical method works here because the life-time of clouds of incoherent excitons is 2-3 orders of magnitude longer than the duration of laser pulses that create them. The resulting repulsive potentials serve as time-dependent perturbations (sort of effective magnetic fields) that enable efficient optical control of the quantum states of the trapped condensate. We believe, the optical control constitutes an important advantage of polariton qubits. Having in mind that the elliptic potentials where polariton condensates are confined are also formed by optical means (with use of a non-resonant cw laser and a spatial-light modulator) we are confident that the scalability of polariton quantum networks may be achieved at a relatively cheap price, as there is no need to grow new structures and use lithography in this case. Last but not least, the operation temperature of superconducting qubits is currently at the milli-Kelvin range. In contrast, the present experiments on polariton quantum beats have been realised at the temperature of about 6K. In the future, passing to the systems that enable polariton superfluidity at the room temperature, such as the perovskite-based microcavities, we hope to realize room-temperature polariton qubits \cite{Kavokin2022}.

\end{widetext}

\end{document}